\documentclass[a4paper,11pt]{article}

\usepackage[normalem]{ulem}
\usepackage{xcolor}
\usepackage[english]{babel}
\usepackage[T1]{fontenc}
\usepackage[utf8]{inputenc}
\usepackage{authblk}
\usepackage{mathtools}
\usepackage{epsfig}
\usepackage{slashed}
\usepackage{amsmath,amssymb}
\usepackage{mathrsfs}
\usepackage{amsfonts}
\usepackage{enumitem}
\usepackage{graphicx,color,xcolor}
\usepackage{cite}
\usepackage{float}
\usepackage{subcaption}
\usepackage{soul}
\usepackage{hyperref}
\usepackage{wrapfig}
\usepackage{booktabs}
\usepackage{array}
\usepackage{microtype}
\usepackage{bm}
\usepackage{cleveref}
\usepackage[left=2.5cm,right=2.5cm,top=2.5cm,bottom=2.5cm]{geometry}
\numberwithin{equation}{section}

\definecolor{refcol}{rgb}{0.9,0.1,0.1}
\hypersetup{colorlinks=true,linkcolor=blue,citecolor=refcol,urlcolor=cyan,linktocpage}
\newcommand{\Tr}{\operatorname{Tr}}
\newcommand{\dd}{\mathrm{d}}
\newcommand{\ii}{\mathrm{i}}
\newcommand{\cD}{\mathcal{D}}
\newcommand{\cC}{\mathcal{C}}

\newcommand{\Thetaf}{\Theta}
\newcommand{\rhozero}{\varrho_{0}}
\newcommand{\mr}{\mathfrak{r}}

\begin{document}

\begin{titlepage}
	\thispagestyle{empty}
	
	\title{{\huge\bf Action-angle variables and phase space formulation of Hermitian matrix models}}
	
	\vfill
	
	\author{
		{\bf Arghya Chattopadhyay}\thanks{{\tt arghya.chattopadhyay@upr.edu}}
		\smallskip\hfill\\
		\small{
			{\it Department of Physics, University of Puerto Rico at Mayag\"uez}\\
			{\it Mayag\"uez, Puerto Rico 00681, USA}
			\vspace{1cm}
		}
	}
	\bigskip\bigskip\bigskip\bigskip
	\vfill
	\date{
		\begin{quote}
			\centerline{{\bf Abstract}}
			{\small
				We develop a phase space description of large $N$ Hermitian one matrix models directly from the recursions satisfied by the orthogonal polynomials. In the one-cut phase, the recursion lattice naturally gives rise to a semiclassical action-angle pair, which can be mapped canonically to eigenvalue and momentum variables. We show that the resulting momentum profile is determined by the density of zeros of the orthogonal polynomials, and that the same phase space structure is reproduced from the Wigner transform of the Christoffel-Darboux projector and from the planar spectral curve. The Gaussian model provides a simple consistency check of the proposal. We then extend the construction to the symmetric quartic two-cut phase, where a period-two Jacobi recursion produces two Bloch bands and two disconnected phase space components with actions given by the partial 't Hooft couplings. Finally, we outline the finite-gap generalization appropriate to multicut phases.
			}
	\end{quote}}
\end{titlepage}

\thispagestyle{empty}
\maketitle
\vfill
\eject

\tableofcontents

\section{Introduction}
\label{sec:introduction}

Hermitian matrix models provide one of the \emph{simplest} settings in which large $N$ quantum field theoretic ideas, random geometry, integrable systems and spectral theory meet naturally. The planar limit underlies the enumeration of random surfaces and two-dimensional gravity \cite{Brezin:1978,DiFrancesco:1993}, while their spectral curves play a central role in topological strings and supersymmetric gauge theories \cite{Dijkgraaf:2002}. The method of orthogonal polynomials gives an independent, nonperturbative description of the same models and exposes their relation to integrable structures.

The phase space and collective field descriptions of matrix quantum mechanics and related large $N$ systems are well known \cite{Jevicki:1980,Das:1990}. In these approaches the main step is the large $N$ reformulation in terms of a continuous eigenvalue density, through which the discrete matrix degrees of freedom reorganize into a field living along the eigenvalue direction. In a series of works \cite{Dutta:2008,Chattopadhyay:2017,Chattopadhyay:2018}, the large $N$ phases of \emph{zero-dimensional} unitary matrix models (UMMs) were described by two dimensional droplets in a \emph{phase space}. The character expansion of the unitary matrix integral is naturally written in terms of Schur polynomials labelled by Young diagrams. At large $N$, the eigenvalue angles provide the position coordinate, while the Young diagram row lengths encode the conjugate momentum, giving a phase space description that combines the eigenvalue and representation bases. Because the eigenvalue coordinate is compact, the momentum is naturally discrete, in direct accord with the integer-valued Young diagram data. This description was subsequently used in the quantization and dynamics of unitary matrix droplets \cite{Chattopadhyay:2021,Chattopadhyay:2022}, as well as used to propose a Hamiltonian for the Riemann zeta function \cite{Chattopadhyay:2018bzs,Dutta:2016byx}. The basic motivation for this paper is to look for a phase space formulation for Hermitian matrix models following the works on UMMs.

The Hermitian one matrix model (HMM) is different in an important way. Its eigenvalues live on the real line, and therefore a conjugate momentum, if a useful phase space formulation exists, should be continuous. Moreover, the model considered here is genuinely zero dimensional, hence there is no time variable and consequently no canonical momentum supplied by a Lagrangian. Nonetheless, the orthogonal polynomial formulation already contains a discrete recursion coordinate, a Jacobi operator, and a fermionic projector. These ingredients allow us to construct a canonical phase space description without introducing any apriori \emph{time direction}. The purpose of this paper is to make this statement precise. For the one-cut phase, we show that the continuum polynomial degree
\begin{equation}
	I=n g_s=t\frac{n}{N},\qquad t=N g_s,
	\label{eq:intro-I}
\end{equation}
defines an action variable, where $n$ labels the orthogonal polynomial, $N$ is the matrix size, and $g_s$ is the matrix model coupling. The Fourier phase $\theta$ associated with the three-term recurrence of the orthogonal polynomials then provides the conjugate angle. A canonical transformation from the action-angle pair $(I,\theta)$ leads to the eigenvalue coordinate $x$ and its continuous conjugate momentum $p$, thereby providing a phase space description of the HMM. The momentum admits three equivalent descriptions. First,
\begin{equation}
	p(x;I)=\pi I\,\rhozero(x;I),
	\label{eq:intro-main}
\end{equation}
where $\rhozero(x;I)$ is the limiting density of zeros of the degree-$n$ orthogonal polynomial. The same momentum defines the boundary of the semiclassical Wigner symbol of the Christoffel-Darboux projector onto the first $n$ orthogonal polynomial states. Finally, for the \emph{outermost curve}, $I=t$, it is related to the boundary value of the planar spectral curve through
\begin{equation}
	p(x;t)=-\frac{\ii}{2}y_{+}(x)=\pi t\,\rho(x),
	\label{eq:intro-spectral}
\end{equation}
where the choice of $y_{+}$ specifies the physical sheet. The polynomial zeros themselves therefore can be interpreted as nodes in coordinate space. Their local spacing reconstructs the momentum according to
\begin{equation}
	p(x_{j,n};I)\simeq
	\frac{\pi g_s}{x_{j+1,n}-x_{j,n}},
	\label{eq:intro-spacing}
\end{equation}
where $x_{j+1,n}$ and $x_{j,n}$ denote neighboring zeros of the degree-$n$ polynomial.

We then show that this construction survives a nontrivial topology change. For the symmetric quartic double well, the two-cut phase is controlled by a slowly varying period-two recurrence \cite{Bleher:1999}. The corresponding $2\times2$ Bloch symbol produces precisely the two matrix model cuts. Its limiting zero density reproduces the known planar density, and the resulting phase space consists of two disconnected droplets. Their individual actions are given by the corresponding partial 't Hooft couplings. For general multicut solutions, the recursion coefficients acquire an oscillatory large $N$ behavior that reflects the distribution of eigenvalues among the different cuts. This behavior is periodic in special cases and quasiperiodic for generic filling fractions \cite{Bonnet:2000}. We distinguish these situations explicitly.

The paper is organized as follows. \Cref{sec:prelim} reviews the planar HMM, the orthogonal polynomial treatment, Jacobi operator and related thongs to fix the conventions used throughout. \Cref{sec:onecut} develops the one-cut construction, by elaborating on the canonical pair $(\theta,I)$, which is mapped to $(x,p)$. The same momentum is then recovered from polynomial zeros, from the Wigner symbol of the projector, and from the planar spectral curve. \Cref{sec:twocut} treats the symmetric quartic two-cut phase and shows how the single site Fourier angle is replaced by the Bloch angle of a two site cell. \Cref{sec:multicut} explains the extension to periodic and generic finite gap multicut recursions, and \cref{sec:conclusion} summarizes the results and open questions. There are four appendices to discuss several technical details and theorems to give a basic pedagogy on Kuijlaars-Van Assche theorem, Christoffel-Darboux kernel and other related structures.

\section{Preliminaries}
\label{sec:prelim}

For completeness, we begin by reviewing the standard matrix model ingredients that will be used in the subsequent sections. In particular, we summarize the planar HMM, its resolvent and spectral curve, the orthogonal polynomial formulation, and the associated determinantal structures. This section is meant to fix notation and collect the necessary background rather than introduce any new results. Our conventions and basic matrix model notation largely follow \cite{Marino:2005}. For a broader modern account the reader may follow \cite{Eynard:2015}.

\subsection{Hermitian one-matrix model and planar curve}

We consider the gauged HMM as
\begin{equation}
	Z_N=\frac{1}{\mathrm{vol}\,U(N)}
	\int \dd M\,
	\exp\left[-\frac{1}{g_s}\Tr W(M)\right],
	\label{eq:matrixZ}
\end{equation}
where $W(M)$ is dubbed as the matrix model potential. After diagonalization,
\begin{equation}
	Z_N=\frac{1}{N!}
	\int\prod_{i=1}^N\frac{\dd x_i}{2\pi}\,
	\Delta^2(x)
	\exp\left[-\frac{1}{g_s}
	\sum_{i=1}^N W(x_i)\right],
	\label{eq:eigenZ}
\end{equation}
where $\Delta(x)$ denotes the Vandermonde determinant\footnote{For the eigenvalues $x_i$ of an $N\times N$ Hermitian matrix,
	\[
	\Delta(x)=\prod_{1\leq i<j\leq N}(x_i-x_j)
	=\det_{1\leq i,j\leq N}\!\left(x_i^{j-1}\right).
	\]
	Its square is the Jacobian associated with diagonalizing the Hermitian matrix.}. We take the 't Hooft limit
\begin{equation}
	N\rightarrow\infty,\qquad
	g_s\rightarrow0,\qquad
	t=N g_s\quad\text{fixed}.
	\label{eq:thooft}
\end{equation}
The eigenvalues condense into a normalized density $\rho(x)$ supported on a set $\cC\subset\mathbb R$, with $\int_{\cC}\rho(x)\dd x=1$. Extremizing the planar effective action generates
\begin{equation}
	\frac{1}{2t}W'(x)
	=\mathcal{P}\int_{\cC}
	\frac{\rho(x')\,\dd x'}{x-x'},
	\qquad x\in\cC,
	\label{eq:saddle}
\end{equation}
where $\mathcal P$ is the Cauchy principal value. The planar resolvent is
\begin{equation}
	\omega_0(z)=\int_{\cC}\frac{\rho(x)\dd x}{z-x},
	\label{eq:resolvent}
\end{equation}
which is analytic away from $\cC$ and behaves as $\omega_0(z)\sim z^{-1}$ as $z\to\infty$. Its discontinuity determines the density,
\begin{equation}
	\rho(x)
	=-\frac{1}{2\pi\ii}
	\lim_{\epsilon\to0^+}
	\left[
	\omega_0(x+\ii\epsilon)-\omega_0(x-\ii\epsilon)
	\right],
	\label{eq:disc}
\end{equation}
and the saddle equation gives the boundary condition
\begin{equation}
	\omega_0(x+\ii\epsilon)+\omega_0(x-\ii\epsilon)
	=\frac{1}{t}W'(x),\qquad x\in\cC.
	\label{eq:RHcondition}
\end{equation}
Therefore solving for the planar resolvent of HMM is equivalent to solving the Riemann-Hilbert problem \eqref{eq:RHcondition}.  For a regular one-cut solution, $\cC\in[a,b]$, and the solution is given by \cite{Migdal:1983qrz} as
\begin{equation}
	\omega_0(p)={1\over 2t}\oint_\mathcal{C}{dz\over 2\pi \ii}{W'(z)\over p-z}\left({(p-a)(p-b)}\over (z-a)(z-b)\right)^{1\over 2}.
	\label{eq:migdal}
\end{equation}
The endpoints $a$ and $b$ are itself fix by imposing the asymptotic condition on the resolvent using the equations
\begin{equation}
	\begin{split}
		\oint_\mathcal{C}{dz\over 2\pi \ii}{W'(z)\over\sqrt{(z-a)(z-b)}}=0\\
		\oint_\mathcal{C}{dz\over 2\pi \ii}{zW'(z)\over\sqrt{(z-a)(z-b)}}=2t.
	\end{split}
	\label{eq:abcond}
\end{equation}
The closed form solution \eqref{eq:migdal} due to \cite{Migdal:1983qrz} is valid for any general polynomial, for a polynomial potential the resolvent gets simplified in the form \cite{Marino:2005}
\begin{equation}
	\omega_0(p)=\frac{1}{2t}
	\left[
	W'(p)-M(p)\sqrt{(p-a)(z-b)}
	\right],
	\label{eq:onecut-resolvent}
\end{equation}
with 
\begin{equation}
	M(p)=\oint_0{dz\over 2\pi \ii}{W'({1/z})\over 1-pz}{1\over \sqrt{(1-az)(1-bz)}},
\end{equation}
where \eqref{eq:abcond} should be used to find the end points. Another way to find the resolvent is to use the spectral function\footnote{For more details on the interpretation of spectral function in terms of topological string theory or certain Calabi-Yau manifold, consult \cite{Marino:2005}.}
\begin{equation}
	y(z)=W'(z)-2t\omega_0(z).
	\label{eq:ydef}
\end{equation}
The values immediately above and below a physical cut are
\begin{equation}
	y_\pm(x)=\lim_{\epsilon\to0^+}y(x\pm\ii\epsilon).
	\label{eq:y-boundary}
\end{equation}
Equations~\eqref{eq:disc} and \eqref{eq:RHcondition} imply
\begin{equation}
	y_+(x)+y_-(x)=0,
	\qquad
	y_+(x)-y_-(x)=4\pi\ii t\rho(x),
\end{equation}
so that
\begin{equation}
		y_+(x)=2\pi\ii t\rho(x),
		\qquad
		y_-(x)=-2\pi\ii t\rho(x).
	\label{eq:y-rho}
\end{equation}
These two are the boundary values of one analytic function on the two sheets of the planar spectral curve.

\subsection{Orthogonal polynomials and the Jacobi operator}
\label{sec:op-prelim}

Introduce monic polynomials $p_n(x)$, orthogonal with respect to the $N$ dependent measure
\begin{equation}
	\dd\mu_N(x)=\frac{\dd x}{2\pi}
	\exp\left[-\frac{N}{t}W(x)\right],
	\qquad
	\int p_n(x)p_m(x)\dd\mu_N(x)=h_n\delta_{nm}.
	\label{eq:opmeasure}
\end{equation}
Because the $p_n$ are monic, the vandermonde can be written as
\begin{equation}
	\Delta(x)=\det_{1\leq i,j\leq N}p_{j-1}(x_i),
\end{equation}
and orthogonality gives
\begin{equation}
	Z_N=\prod_{n=0}^{N-1}h_n
	=h_0^N\prod_{n=1}^{N-1}r_n^{N-n},
	\qquad
	r_n=\frac{h_n}{h_{n-1}}.
	\label{eq:Zhn}
\end{equation}
The monic three-term recurrence relation is then\footnote{For an even potential, parity gives $s_n=0$.}
\begin{equation}
	xp_n(x)=p_{n+1}(x)-s_n p_n(x)+r_n p_{n-1}(x).
	\label{eq:monic-recurrence}
\end{equation}
For the orthonormal polynomials $P_n=p_n/\sqrt{h_n}$ this becomes
\begin{equation}
	xP_n(x)=-s_nP_n(x)
	+\sqrt{r_{n+1}}\,P_{n+1}(x)
	+\sqrt{r_n}\,P_{n-1}(x).
	\label{eq:recurrence}
\end{equation}
Equation~\eqref{eq:recurrence} is the spectral equation for the Jacobi matrix
\begin{equation}
	J=
	\begin{pmatrix}
		-s_0 & \sqrt{r_1} & 0 & \cdots\\
		\sqrt{r_1} & -s_1 & \sqrt{r_2} & \cdots\\
		0 & \sqrt{r_2} & -s_2 & \cdots\\
		\vdots & \vdots & \vdots & \ddots
	\end{pmatrix},
	\qquad
	J\mathbf P(x)=x\mathbf P(x),
	\label{eq:jacobi-spectral}
\end{equation}
with $\mathbf P=(P_0,P_1,\ldots)^T$. Thus multiplication by the matrix eigenvalue coordinate is represented in the orthogonal polynomial basis by the Jacobi operator. The recurrence formulation and its relation to Toda-type integrable structures are standard in the literature \cite{DiFrancesco:1993,Adler:1995,Takasaki:1995}.

Let $J_n$ be the $n\times n$ principal truncation of $J$. The determinant $D_n(x)=\det(x\mathbf 1_n-J_n)$ satisfies the same monic recurrence and initial data as $p_n(x)$, hence
\begin{equation}
p_n(x)=\det(x\mathbf 1_n-J_n).
	\label{eq:zeros-jacobi}
\end{equation}
Therefore the $n$ zeros of $p_n$ are exactly the eigenvalues of $J_n$. Their large $n$ distribution will play the central role in this paper.

\subsection{Fermionic projector}
\label{sec:projector-prelim}

Define the normalized functions
\begin{equation}
	\psi_n(x)=\frac{1}{\sqrt{2\pi}}
	e^{-W(x)/(2g_s)}P_n(x),
	\qquad
	\int_{\mathbb R}\dd x\,\psi_n(x)\psi_m(x)=\delta_{nm}.
	\label{eq:wavefunctions}
\end{equation}
The Vandermonde determinant then rewrites the normalized joint eigenvalue distribution as a Slater determinant,
\begin{equation}
	\mathsf P_N(x_1,\ldots,x_N)
	=\frac{1}{N!}
	\left|
	\det_{1\leq i,j\leq N}\psi_{j-1}(x_i)
	\right|^2.
	\label{eq:slater}
\end{equation}
Thus HMM defines the same determinantal position-space measure as $N$ fermions occupying the one-particle states $\psi_0,\ldots,\psi_{N-1}$ \cite{Ginsparg:1993is,Dijkgraaf:2004}. More generally, the first $n$ states define the rank-$n$ projector
\begin{equation}
	\widehat\Pi_n=\sum_{k=0}^{n-1}|\psi_k\rangle\langle\psi_k|,
	\qquad
	K_n(x,x')=\langle x|\widehat\Pi_n|x'\rangle
	=\sum_{k=0}^{n-1}\psi_k(x)\psi_k(x').
	\label{eq:kernel}
\end{equation}
Orthogonality gives
\begin{equation}
	\int\dd z\,K_n(x,z)K_n(z,x')=K_n(x,x'),
	\qquad
	\int\dd x\,K_n(x,x)=n,
	\label{eq:projector-properties}
\end{equation}
so that $\widehat{\Pi}_n^2=\widehat{\Pi}_n$ and its rank is $n$. The kernel is the weighted Christoffel-Darboux kernel and is the integral kernel of the orthogonal projection onto the first $n$ states \cite{Simon:2008}. Its determinantal meaning and the universal sine-kernel bulk\footnote{Away from the spectral edges, where the limiting density remains nonzero} limit used later are reviewed in appendix \ref{app:cd}. The free fermion interpretation of the matrix model eigenvalue measure arising from the same orthogonal polynomial construction is standard \cite{Ginsparg:1993is} and  related fermionic and geometric interpretations of matrix model wavefunctions have been discussed in \cite{Dijkgraaf:2004}.

In the following we use this standard projector in a different way. Since $\widehat{\Pi}_n$ is a well defined finite $n$ one particle operator, we can associate with it a Wigner symbol using the eigenvalue coordinate $x$, without introducing a time direction or an auxiliary matrix quantum mechanics. We will show that its semiclassical support coincides with the phase space region obtained independently from the action-angle formulation.

\section{The one-cut phase}
\label{sec:onecut}

\subsection{From the recursion lattice to a canonical pair}
\label{sec:canonical-pair}

For one-cut solution, the standard large $N$ orthogonal polynomial analysis assumes the smooth limits
\begin{equation}
	r_n\longrightarrow R(\xi),
	\qquad
	s_n\longrightarrow s(\xi),
	\qquad
	\xi=\frac{n}{N}\in(0,1],
	\label{eq:smoothrec}
\end{equation}
along with,
\begin{equation}
	r_{n+1}=R\!\left(\xi+\frac1N\right)
	=R(\xi)+O(N^{-1}).
\end{equation}
Implying that the recurrence coefficients vary slowly between neighboring sites. In the 't Hooft limit the natural dimensionful coordinate is
\begin{equation}
I=g_s n=t\xi,
	\label{eq:Idef}
\end{equation}
whose \emph{lattice spacing} is $g_s$. Thus $I$ becomes continuous on $0\leq I\leq t$ when $g_s\to0$. To identify its conjugate variable, regard $n$ as the coordinate of a discrete recursion lattice. Introduce an abstract basis $|n\rangle$ and\footnote{The ket notation is only a convenient representation of the Jacobi recursion. It does not introduce any additional physical dynamical system.}
\begin{equation}
	\hat n|n\rangle=n|n\rangle,
	\qquad
	\hat I=g_s\hat n,
	\label{eq:Ioperator}
\end{equation}
together with the one-site shift
\begin{equation}
	U|n\rangle=|n+1\rangle.
	\label{eq:shift}
\end{equation}
Then
\begin{equation}
	[\hat I,U]=g_sU.
	\label{eq:commIU}
\end{equation}
Locally away from the endpoint $n=0$, the lattice admits a Fourier representation. With
\begin{equation}
	|\theta\rangle\sim\sum_n e^{-\ii n\theta}|n\rangle,
	\qquad \theta\sim\theta+2\pi,
\end{equation}
one finds
\begin{equation}
	U\longrightarrow e^{\ii\theta},
	\qquad
	U^{-1}\longrightarrow e^{-\ii\theta}.
	\label{eq:Usymbol}
\end{equation}
The leading, or principal, symbol\footnote{By ``symbol'' we mean the
	ordinary function obtained in the semiclassical limit by replacing the
	recursion operators by their continuum variables,
	\begin{equation}
		\hat I\longrightarrow I,
		\qquad
		U\longrightarrow e^{\ii\theta}.
\end{equation}}
retains the $g_s\to0$ part of this correspondence.  The leading
commutator of two operators induces a Poisson bracket on their
principal symbols through
\begin{equation}
	{\rm symb}\,[\hat A,\hat B]
	=
	\ii g_s\{A,B\}
	+O(g_s^2).
	\label{eq:semicomm}
\end{equation}
At this stage the normalization of the Poisson bracket between $I$
and $\theta$ has not yet been specified; it will follow from the
recursion algebra itself.

Applying \eqref{eq:semicomm} to the exact shift relation
\eqref{eq:commIU},
\begin{equation}
	[\hat I,U]=g_s U,
\end{equation}
and using $U\rightarrow e^{\ii\theta}$ gives
\begin{equation}
	\ii g_s\{I,e^{\ii\theta}\}
	=
	g_s e^{\ii\theta},
\end{equation}
or
\begin{equation}
	\{I,e^{\ii\theta}\}
	=
	-\ii e^{\ii\theta}.
\end{equation}
Since a Poisson bracket acts as a derivation,
\begin{equation}
	\{I,e^{\ii\theta}\}
	=
	\ii e^{\ii\theta}\{I,\theta\},
\end{equation}
we obtain
\begin{equation}
	\boxed{\{\theta,I\}=1.}
	\label{eq:canonicalthetaI}
\end{equation}
Thus the continuum recursion coordinate $I$ and the Fourier phase
$\theta$ form a canonically normalized pair. We use
\begin{equation}
	\{f,g\}_{\theta,I}
	=\partial_\theta f\,\partial_I g
	-\partial_I f\,\partial_\theta g,
	\qquad
	\Omega=\dd\theta\wedge\dd I.
	\label{eq:PBthetaI}
\end{equation}
The meaning of $I$ as an action, rather than merely as a canonical coordinate, will be fixed by the phase space area discussed in \cref{sec:wigner-main}.

We now apply the same semiclassical correspondence to the Jacobi
operator itself.  At a fixed continuum level $I$, the recurrence
coefficients vary only by $O(N^{-1})$ between neighboring sites and
may therefore be treated locally as constants.  Defining
\begin{equation}
	a(I)=\sqrt{R(I/t)},
	\qquad
	b(I)=-s(I/t),
	\label{eq:abdef}
\end{equation}
the three-term recurrence relation takes the local form
\begin{equation}
	xP_n
	\simeq
	b(I)P_n
	+a(I)P_{n+1}
	+a(I)P_{n-1}.
	\label{eq:local-recurrence}
\end{equation}
Using the shift operator introduced in Eq.~\eqref{eq:shift}, this may
equivalently be written as
\begin{equation}
	J_{\rm loc}(I)
	=
	b(I)+a(I)U+a(I)U^{-1},
	\label{eq:local-jacobi-operator}
\end{equation}
where $J_{\rm loc}(I)$ denotes the Jacobi operator with its recurrence
coefficients frozen at the level $I$. The Fourier representation of the shift operator was already obtained in \eqref{eq:Usymbol}, implying the leading semiclassical symbol of the locally frozen Jacobi operator as
\begin{align}
	X(I,\theta)
	&=
	b(I)
	+a(I)e^{\ii\theta}
	+a(I)e^{-\ii\theta}
	\nonumber\\
	&=
	b(I)+2a(I)\cos\theta .
	\label{eq:jacobi-symbol-intermediate}
\end{align}
Here $X(I,\theta)$ is the ordinary function obtained from the Jacobi operator after replacing the shift operators by their Fourier eigenvalues. Going back to the spectral equation \eqref{eq:jacobi-spectral}, we know that  the spectral value of the Jacobi operator is precisely the same variable $x$ that appears as the eigenvalue coordinate of the matrix model. We therefore obtain
\begin{equation}
	x=X(I,\theta)
	=
	b(I)+2a(I)\cos\theta .
	\label{eq:jacobisymbol}
\end{equation}
At fixed $I$, varying $\theta$ over one period gives the local spectral
band
\begin{equation}
	b(I)-2a(I)
	\leq x\leq
	b(I)+2a(I).
	\label{eq:localband}
\end{equation}
Because the recurrence coefficients vary only by $O(N^{-1})$ between neighboring sites, they may be treated as constant over a local region of the recursion lattice. Around a fixed continuum level $I$ the three-term recurrence therefore reduces, at leading order in $1/N$, to
\begin{equation}
	x\,\phi_n
	=
	b(I)\phi_n
	+a(I)\phi_{n+1}
	+a(I)\phi_{n-1},
	\label{eq:local-jacobi}
\end{equation}
Where $\phi_n$ describes the local dependence on the recursion index and should not be confused with the full orthogonal polynomial $P_{n}(x)$. Equation \ref{eq:local-jacobi} contains no explicit dependence on $n$ and is therefore invariant under translations along the local recursion lattice. Its elementary modes can consequently be chosen as eigenfunctions of the one-site shift operator $U$ as,
\begin{equation}
	\phi_n(\theta)\propto e^{\ii n\theta},
	\qquad
	\phi_{n\pm1}(\theta)
	=
	e^{\pm\ii\theta}\phi_n(\theta).
\end{equation}
Substitution this into \eqref{eq:local-jacobi} gives the local dispersion relation we get back
\begin{equation}
	x=b(I)+2a(I)\cos\theta .
	\label{eq:local-dispersion}
\end{equation}
This reproduces precisely the leading symbol $X(I,\theta)$ of the Jacobi operator obtained from the shift-operator construction above\footnote{The structure is analogous to a one dimensional nearest neighbor tight binding model, whose local dispersion relation takes the form $E(k)=E_0+2\kappa\cos k$. Here the polynomial degree $n$ plays the role of the lattice site, $\theta$ is the corresponding Fourier or Bloch variable, and the matrix model eigenvalue $x$ plays the role of the spectral value.}. The significance of this agreement is that the same Fourier variable $\theta$ that is canonically conjugate to the continuum recursion coordinate $I$ also parameterizes the local spectrum of the Jacobi operator. The recurrence relation therefore supplies an explicit map from the canonical variables $(I,\theta)$ to the matrix model eigenvalue coordinate \eqref{eq:jacobisymbol}. Thus $x$ may be regarded as a coordinate function on the $(I,\theta)$ phase space. For fixed $I$, varying $\theta$ over one period traces the range of eigenvalue coordinates accessible at that recursion level \eqref{eq:localband}, which is the local spectral band of the frozen Jacobi operator. Further, for a point $x$ in the interior of this band, \eqref{eq:jacobisymbol} has two angular branches,
\begin{equation}
	\theta_\pm(x,I)
	=
	\mp\arccos\!\left(
	\frac{x-b(I)}{2a(I)}
	\right),
	\label{eq:theta-branches}
\end{equation}
corresponding to the two sides of the constant-$I$ trajectory. This two-valued relation is the starting point for constructing a momentum $p_\pm(x;I)$ such that the transformation $(\theta,I)\rightarrow(x,p)$ is canonical, which will be elaborated in the next subsection.

This same parameterization also determines the local density of states. The Fourier variable carries the uniform measure $\dd\theta/(2\pi)$. We want the corresponding density in $x$, given $x=b+2a\cos\theta$. Summing the contribution of the two branches in \eqref{eq:theta-branches} therefore gives
\begin{align}
	\sigma_I(x)
	&=
	\frac{1}{2\pi}
	\sum_{\theta:\,X(I,\theta)=x}
	\left|
	\frac{\dd\theta}{\dd x}
	\right|
	\nonumber\\
	&=
	\frac{1}{\pi}
	\frac{
		\Thetaf\!\left(4a^2(I)-[x-b(I)]^2\right)
	}{
		\sqrt{4a^2(I)-[x-b(I)]^2}
	}.
	\label{eq:localshell}
\end{align}
where the $\Thetaf$ function is there to make it evident that the density vanishes outside this band. This is the normalized density associated with the locally constant Jacobi operator, $\int\dd x\,\sigma_I(x)=1$.
The quantity $\sigma_I(x)$ is a local density associated with a single recursion level $I$. It should not yet be identified with the planar matrix model eigenvalue density. The latter is obtained by averaging the local spectral measures over the continuum of recursion levels,
\begin{equation}
	\rho(x)
	=
	\frac{1}{t}\int_0^t\dd I\,\sigma_I(x)
	=
	\int_0^1\dd\xi\,\sigma_{t\xi}(x).
	\label{eq:rho-local-average}
\end{equation}
Using $a^2(I)=R(I/t)$ and $b(I)=-s(I/t)$, this becomes
\begin{equation}
	\rho(x)
	=
	\int_0^1\frac{\dd\xi}{\pi}\,
	\frac{
		\Thetaf\!\left(
		4R(\xi)-[x+s(\xi)]^2
		\right)
	}{
		\sqrt{
			4R(\xi)-[x+s(\xi)]^2
		}
	},
	\label{eq:rho-op}
\end{equation}
which is the standard large $N$ orthogonal polynomial expression for the planar density \cite{Marino:2005}. What is new for our purposes is that the Fourier variable $\theta$ has already been identified above as the canonical variable conjugate to $I=g_sn$.

\subsection{Canonical transformation to \texorpdfstring{$(x,p)$}{(x,p)}}
\label{sec:canonical-xp}

Equation \eqref{eq:jacobisymbol} gives the matrix eigenvalue coordinate $x$ as a function of the canonical variables $(\theta,I)$. To obtain a phase space description in which $x$ itself is used as the coordinate, we must construct a conjugate variable $p$ such that the transformation
\begin{equation}
	(\theta,I)\longrightarrow(x,p)
\end{equation}
is canonical. The transformation may be constructed locally using a
generating function $S_\pm(x,I)$.  The relation $x=X(I,\theta)$ may be inverted locally, on either
branch of the spectral band, to give $\theta=\theta_\pm(x,I)$.  We
now seek a generating function whose independent variables are $x$
and $I$.  Since the transformation is required to preserve
\begin{equation}
	d\theta\wedge dI=dx\wedge dp ,
\end{equation}
the one-form $p\,dx+\theta\,dI$ is locally closed and may therefore be
written as
\begin{equation}
	dS_\pm=p_\pm\,dx+\theta_\pm\,dI .
	\label{eq:dS}
\end{equation}
Thus
\begin{equation}
	p_\pm=
	\left.\frac{\partial S_\pm}{\partial x}\right|_I,
	\qquad
	\theta_\pm=
	\left.\frac{\partial S_\pm}{\partial I}\right|_x .
	\label{eq:generating-relations}
\end{equation}
Since $\theta_\pm(x,I)$ is already known, the second relation can be
integrated with respect to $I$ at fixed $x$.  The most general local
solution is therefore
\begin{equation}
	S_\pm(x,I)
	=
	\int_{I_0}^{I}
	\theta_\pm(x,J)\,\dd J
	+
	F_\pm(x),
	\label{eq:generating}
\end{equation}
where $I_0$ is an arbitrary reference level. The function $F_\pm(x)$ represents the remaining freedom in the choice of the momentum origin.  Equality of mixed derivatives of $S$ therefore implies
\begin{equation}
		\left.
		\frac{\partial p}{\partial I}
		\right|_x
		=
		\left.
		\frac{\partial\theta}{\partial x}
		\right|_I .
	\label{eq:mixed}
\end{equation}
We now regard $x$ as the function $x=x(\theta,I)$ obtained from the
Jacobi symbol.  Accordingly,
\begin{equation}
	p(\theta,I)
	=
	p\bigl(x(\theta,I),I\bigr),
\end{equation}
and the derivatives entering the Poisson bracket must be evaluated
using the chain rule.  At fixed $\theta$,
\begin{equation}
	\left.
	\frac{\partial p}{\partial I}
	\right|_\theta
	=
	\left.
	\frac{\partial p}{\partial I}
	\right|_x
	+
	\left.
	\frac{\partial p}{\partial x}
	\right|_I
	\left.
	\frac{\partial x}{\partial I}
	\right|_\theta ,
	\label{eq:chain-I}
\end{equation}
whereas at fixed $I$,
\begin{equation}
	\left.
	\frac{\partial p}{\partial\theta}
	\right|_I
	=
	\left.
	\frac{\partial p}{\partial x}
	\right|_I
	\left.
	\frac{\partial x}{\partial\theta}
	\right|_I .
	\label{eq:chain-theta}
\end{equation}
Using the Poisson bracket in the original canonical variables and using \eqref{eq:chain-I} and \eqref{eq:chain-theta}
\begin{align}
	\{x,p\}_{\theta,I}
	&=
	\left.
	\frac{\partial x}{\partial\theta}
	\right|_I
	\left.
	\frac{\partial p}{\partial I}
	\right|_\theta
	-
	\left.
	\frac{\partial x}{\partial I}
	\right|_\theta
	\left.
	\frac{\partial p}{\partial\theta}
	\right|_I= 	\left.
	\frac{\partial x}{\partial\theta}
	\right|_I	\left.
	\frac{\partial p}{\partial I}
	\right|_x
\end{align}
Using \eqref{eq:mixed} then gives
\begin{equation}
	\{x,p\}_{\theta,I}
	=\left.
	\frac{\partial x}{\partial\theta}
	\right|_I 	\left.
	\frac{\partial\theta}{\partial x}
	\right|_I =1.
\end{equation}
Thus $(x,p)$ is a canonical pair, as already implied by the equality
of symplectic forms. The momentum can now be obtained explicitly. Differentiating \eqref{eq:theta-branches} at fixed $I$ gives, inside the local spectral band,
\begin{equation}
	\left.
	\frac{\partial\theta_\pm}{\partial x}
	\right|_I
	=
	\pm
	\frac{1}{
		\sqrt{
			4a^2(I)-[x-b(I)]^2
		}
	}.
	\label{eq:thetax}
\end{equation}
Using \eqref{eq:mixed}, this is equivalently
\begin{equation}
	\left.
	\frac{\partial p_\pm}{\partial I}
	\right|_x
	=
	\pm
	\frac{1}{
		\sqrt{
			4a^2(I)-[x-b(I)]^2
		}
	}.
	\label{eq:pI}
\end{equation}
Note that for a fixed $x$, a given value of $J$ contributes only when $x$ lies inside the corresponding local Jacobi band,
\begin{equation}
	4a^2(J)-[x-b(J)]^2\geq0.
\end{equation}
Equation \eqref{eq:pI} determines the $I$-dependence of the momentum, but leaves an
$x$-dependent integration freedom. The general local solution for \eqref{eq:pI} is
\begin{equation}
	p_\pm(x;I)
	=
	\pm\int \dd I\,
	\frac{1}{
		\sqrt{4a^2(I)-[x-b(I)]^2}}
	+
	C_\pm(x),
\end{equation}
Equivalently, choosing a reference recursion level $I_0$,
\begin{equation}
	p_\pm(x;I)-p_\pm(x;I_0)
	=
	\pm\int_{I_0}^{I}\dd J\,
	\frac{1}{
		\sqrt{4a^2(J)-[x-b(J)]^2}},
\end{equation}
or
\begin{equation}
	p_\pm(x;I)
	=
	\pm\int_{I_0}^{I}\dd J\,
	\frac{1}{
		\sqrt{4a^2(J)-[x-b(J)]^2}}
	+
	C_\pm(x).
\end{equation}
The function $C_\pm(x)$ is precisely the freedom represented by
$F_\pm'(x)$ in the generating function.  This freedom cannot be fixed by canonicality alone as a transformation $p\rightarrow p+f(x)$ leaves $dx\wedge dp$ unchanged. Thus the canonical construction determines a family of symplectically equivalent momentum coordinates. The orthogonal polynomial problem itself, however, provides a natural normalization. In the next subsection we show that the limiting distribution of polynomial zeros selects a distinguished representative of this family.

\subsection{Zeros of orthogonal polynomials and momentum}
\label{sec:zeros-momentum}

Let $x_{1,n},\ldots,x_{n,n}$ be the real zeros of the monic
degree-$n$ polynomial $p_n(x)$ and define the normalized zero-counting
measure
\begin{equation}
	\dd\nu_n(x)
	=
	\frac{1}{n}
	\sum_{j=1}^{n}
	\delta(x-x_{j,n})\,\dd x .
	\label{eq:zero-counting}
\end{equation}
By Eq.~\eqref{eq:zeros-jacobi}, these zeros are precisely the eigenvalues of the $n\times n$ principal truncation $J_n$ of the Jacobi operator. Their large-$n$ distribution is therefore controlled by the same recurrence coefficients that entered the canonical construction above.

The Theorem due to Kuijlaars and Van Assche \cite{Kuijlaars:1999}, stated in detail and translated to our conventions in Appendix \ref{app:kva}, determines this limiting distribution. Under the smooth scaling assumption \eqref{eq:smoothrec}, the normalized
zero-counting measures converge weakly, for $n,N\to\infty$ with $I=ng_s=t\,n/N$ fixed, to a density
\begin{equation}
		\rhozero(x;I)
		=
		\frac{1}{I}
		\int_0^I\dd I'\,\sigma_{I'}(x) .
	\label{eq:zerodensity}
\end{equation}
Here $\sigma_{I'}(x)$ is the local spectral density of the Jacobi operator with its recurrence coefficients frozen at the recursion level $I'$, introduced in \eqref{eq:localshell}. This zero-density theorem now identifies a particularly natural
representative of the canonical family discussed above. Indeed, \eqref{eq:pI} and the definition of the local density give, wherever $x$ lies inside the local Jacobi band,
\begin{equation}
	\left.
	\frac{\partial p_\pm}{\partial I}
	\right|_x
	=
	\pm\pi\,\sigma_I(x).
	\label{eq:pI-sigma}
\end{equation}
On the other hand, \eqref{eq:zerodensity} implies
\begin{equation}
	\frac{\partial}{\partial I}
	\left[
	I\,\rhozero(x;I)
	\right]
	=
	\sigma_I(x).
	\label{eq:zero-derivative}
\end{equation}
Thus $\pi I\rhozero(x;I)$ has precisely the $I$-dependence required of the canonical momentum. The matrix model zero distribution therefore selects the distinguished global representative
\begin{equation}
		p_\pm(x;I)
		=
		\pm\pi I\,\rhozero(x;I)
		=
		\pm\pi
		\int_0^I\dd I'\,\sigma_{I'}(x).
	\label{eq:pzeros}
\end{equation}
In terms of the generating function of the previous subsection, this
choice fixes the otherwise arbitrary $x$-dependent shift
$p_\pm\rightarrow p_\pm+F_\pm'(x)$.  Rather than canonicality, \eqref{eq:pzeros} fixes this freedom by using the spectral data of the matrix model.

The zeros themselves should therefore not be interpreted as momentum
eigenvalues.  Rather, the polynomial degree $n$, through
$I=ng_s$, specifies the continuum recursion level, while the spatial
distribution of the zeros determines the magnitude of the conjugate
momentum on the corresponding constant-$I$ curve.  In this sense the
zero distribution provides a coordinate-space reconstruction of the
phase space profile.

This statement also has a simple local interpretation.  Let
$x_{j,n}$ and $x_{j+1,n}$ be neighboring zeros in a regular bulk
region, where the limiting density varies slowly on the scale of one
zero spacing. Since $\rhozero$ is normalized to unity, the expected
number of zeros in a small interval $\dd x$ is
$n\rhozero(x;I)\dd x$.  Consequently,
\begin{equation}
	n\,\rhozero(x_{j,n};I)
	\left(
	x_{j+1,n}-x_{j,n}
	\right)
	\simeq 1 .
	\label{eq:zero-spacing-density}
\end{equation}
Using $I=ng_s$ together with Eq.~\eqref{eq:pzeros} gives
\begin{equation}
		|p(x_{j,n};I)|
		\simeq
		\frac{\pi g_s}{
			x_{j+1,n}-x_{j,n}}
		.
	\label{eq:spacing}
\end{equation}
Thus closely spaced zeros correspond to large momentum, whereas a larger nodal separation corresponds to smaller momentum. This is the familiar semiclassical inverse relation between local nodal spacing and momentum, here obtained directly from the orthogonal polynomial recursion and its limiting zero distribution. Note that \eqref{eq:spacing} is a bulk relation, away from the endpoints where the limiting density remains nonzero. A particularly important case is the outermost polynomial, $n=N$.
Then $I=t$, and Eq.~\eqref{eq:zerodensity} becomes
\begin{equation}
	\rhozero(x;t)
	=
	\frac{1}{t}
	\int_0^t\dd I'\,\sigma_{I'}(x).
\end{equation}
This is exactly the orthogonal polynomial representation \eqref{eq:rho-local-average} of the planar matrix-model eigenvalue
density.  Hence
\begin{equation}
		\rhozero(x;t)=\rho(x).
	\label{eq:outer-zero-density}
\end{equation}
The outermost momentum branches are therefore
\begin{equation}
		p_\pm(x;t)
		=
		\pm\pi t\,\rho(x).
	\label{eq:pF}
\end{equation}
This result may finally be compared with the planar spectral curve.
As shown in \eqref{eq:y-rho}, the two boundary values of the
spectral function on the physical cut satisfy
\begin{equation}
	y_\pm(x)
	=
	\pm2\pi\ii t\,\rho(x).
\end{equation}
Combining this relation with Eq.~\eqref{eq:pF} gives
\begin{equation}
		p_\pm(x;t)
		=
		-\frac{\ii}{2}\,y_\pm(x)
		=
		\pm\pi t\,\rho(x).
	\label{eq:p-y}
\end{equation}
Thus the two boundary values of the planar spectral function reproduce
the upper and lower boundaries of the outermost real phase space
curve. Equivalently, the planar spectral curve provides a complex
continuation of the real phase space boundary obtained from the
orthogonal polynomial construction.

\subsection{Wigner transform of the orthogonal polynomial projector}
\label{sec:wigner-main}

The preceding subsections constructed a phase space momentum from the Jacobi recursion and showed that its magnitude is fixed by the limiting density of zeros of the orthogonal polynomials,
\begin{equation}
	p_\pm(x;I)
	=
	\pm\pi I\,\rhozero(x;I).
\end{equation}
We also found a second and conceptually independent route to the same phase space geometry. It starts directly from the finite rank one-particle projector introduced in section \ref{sec:prelim}. This provides an important check of the construction because no canonical momentum is assumed at the outset.

We begin by emphasising that the coordinate-space kernel $K_n(x_1,x_2)=\langle x_1|\widehat{\Pi}_n|x_2\rangle$ is bilocal. The natural variables for passing from such a bilocal operator to a phase space description are its center and relative coordinates,
\begin{equation}
	x=\frac{x_1+x_2}{2},
	\qquad
	\eta=x_1-x_2.
	\label{eq:center-relative}
\end{equation}
The center coordinate $x$ is precisely the matrix-model eigenvalue coordinate used throughout the previous construction. The relative coordinate $\eta$ instead measures the separation between the two arguments of the kernel. Fourier transforming with respect to this relative separation $\eta$ introduces a conjugate variable, which we denote
by $p$. This is the standard Weyl-Wigner map from an operator to a function on phase space. Since $g_s$ is the semiclassical parameter of the matrix model, in direct analogy with $\hbar$ in ordinary quantum mechanics, we define the Weyl symbol of $\widehat{\Pi}_n$ by
\begin{equation}
	u_n(x,p)
	=
	\int_{-\infty}^{\infty}\dd\eta\,
	e^{-\ii p\eta/g_s}
	K_n\!\left(
	x+\frac{\eta}{2},
	x-\frac{\eta}{2}
	\right).
	\label{eq:wigner}
\end{equation}
At this stage the variable $p$ has only one meaning: it is the Fourier variable conjugate to the relative coordinate $\eta$. In particular, we have not identified it with the momentum obtained from the canonical transformation in section\ref{sec:canonical-xp}. If the two constructions lead to the same $p$, this will therefore be a result rather than an assumption.

The normalization in \eqref{eq:wigner} is chosen so that the inverse transform is
\begin{equation}
	K_n\!\left(
	x+\frac{\eta}{2},
	x-\frac{\eta}{2}
	\right)
	=
	\int\frac{\dd p}{2\pi g_s}\,
	e^{\ii p\eta/g_s}\,
	u_n(x,p).
	\label{eq:wigner-inverse-main}
\end{equation}
Setting $\eta=0$ gives
\begin{equation}
	K_n(x,x)
	=
	\int\frac{\dd p}{2\pi g_s}\,
	u_n(x,p),
	\label{eq:wigner-diagonal}
\end{equation}
and integrating over $x$ therefore yields the usual Weyl trace formula
\begin{equation}
	\Tr\widehat{\Pi}_n
	=
	\int
	\frac{\dd x\,\dd p}{2\pi g_s}\,
	u_n(x,p).
	\label{eq:wigner-trace}
\end{equation}
At finite $n$, $u_n(x,p)$ is a quantum phase space symbol and should not be interpreted as an ordinary positive probability distribution. Nevertheless, it retains an exact memory of the fact that $\widehat{\Pi}_n$ is a projector. Since
\begin{equation}
	\widehat{\Pi}_n^{\,2}
	=
	\widehat{\Pi}_n,
\end{equation}
operator multiplication becomes, under the Weyl map,
\begin{equation}
	u_n\star u_n=u_n,
	\label{eq:star-projector}
\end{equation}
where $\star$ denotes the Moyal product. The detailed Weyl-transform conventions and the corresponding Moyal-product formulas are collected in appendix \ref{app:wigner}. We now take the same semiclassical scaling as in the recursion analysis,
\begin{equation}
	n\rightarrow\infty,
	\qquad
	g_s\rightarrow0,
	\qquad
	I=ng_s
	\quad\text{fixed}.
	\label{eq:wigner-scaling}
\end{equation}
Away from regions where the symbol changes on the microscopic $g_s$ scale, the Moyal product reduces at leading order to the ordinary product.  Equation \eqref{eq:star-projector} then reduces to
\begin{equation}
	u_I^2(x,p)=u_I(x,p).
\end{equation}
The smooth leading-order solutions are therefore $u_I=0$ and $u_I=1$. Thus a finite-rank quantum projector becomes, in the semiclassical limit, an occupied or unoccupied region of phase space:
\begin{equation}
	u_I(x,p)
	\simeq
	\begin{cases}
		1, & (x,p)\in\mathcal D_I,\\[1mm]
		0, & (x,p)\notin\mathcal D_I.
	\end{cases}
	\label{eq:droplet-characteristic}
\end{equation}
The transition between these two values is smoothed at finite $g_s$ near the boundary of $\mathcal D_I$, where the derivative expansion of the Moyal product is no longer uniform\footnote{This projector-to-droplet limit is a standard feature of semiclassical Weyl calculus.Closely related rigorous results are discussed in \cite{Cunden:2025}.}.

The projector property therefore tells us that a phase space droplet must arise, but it does not yet determine its shape.  That information is contained in the local structure of the Christoffel-Darboux kernel. As reviewed in appendix \ref{app:cd}, in a regular bulk
region which is away from spectral edges, where the limiting density remains nonzero and varies slowly on the microscopic scale, the kernel takes the universal sine-kernel form
\begin{equation}
	K_n\!\left(
	x+\frac{\eta}{2},
	x-\frac{\eta}{2}
	\right)
	\simeq
	\frac{
		\sin\!\left[
		\pi n\rhozero(x;I)\eta
		\right]
	}{
		\pi\eta
	}.
	\label{eq:sinekernel}
\end{equation}
The appearance of the combination $n\rhozero(x;I)$ has a simple meaning. Since $\rhozero(x;I)$ is normalized to unity and $P_n$ has $n$ zeros, the local number of zeros per unit length is $n\rhozero(x;I)$. The corresponding microscopic spacing is therefore
\begin{equation}
	\Delta x
	\sim
	\frac{1}{n\rhozero(x;I)}.
	\label{eq:microscopic-spacing}
\end{equation}
The sine kernel hence resolves separations $\eta$ precisely on the scale of the local zero spacing. It should be noted that the approximation \eqref{eq:sinekernel} controls separations of the order of the \emph{microscopic spacing}, $\eta=O(1/n)$, rather than providing a uniform approximation to the kernel for arbitrary $\eta$. In what follows we use it to extract the leading local semiclassical Wigner symbol in the bulk.

We may now determine the droplet boundary explicitly. At leading local semiclassical order, substituting \eqref{eq:sinekernel} into the Wigner transform \eqref{eq:wigner} gives
\begin{equation}
	u_n(x,p)
	\simeq
	\int_{-\infty}^{\infty}
	\frac{\dd\eta}{\pi\eta}\,
	e^{-\ii p\eta/g_s}
	\sin\!\left[
	\pi n\rhozero(x;I)\eta
	\right].
	\label{eq:wigner-sine}
\end{equation}
The required Fourier transform is elementary. For $A>0$,
\begin{equation}
	\frac{\sin(A\eta)}{\pi\eta}
	=
	\frac{1}{2\pi}
	\int_{-A}^{A}\dd q\,
	e^{\ii q\eta},
	\label{eq:sinc-box}
\end{equation}
so its Fourier transform is the characteristic function of the
interval $[-A,A]$.  Equivalently,
\begin{equation}
	\int_{-\infty}^{\infty}
	\frac{\dd\eta}{\pi\eta}\,
	e^{-\ii k\eta}\sin(A\eta)
	=
	\Thetaf(A-|k|),
	\label{eq:sinc-fourier}
\end{equation}
up to the conventional half-value at the endpoints.  In the present
case,
\begin{equation}
	k=\frac{p}{g_s},
	\qquad
	A=\pi n\rhozero(x;I).
\end{equation}
The Wigner symbol is therefore nonzero whenever
\begin{equation}
	\frac{|p|}{g_s}
	<
	\pi n\rhozero(x;I).
\end{equation}
Using $I=ng_s$, this condition becomes
\begin{equation}
	|p|
	<
	\pi I\rhozero(x;I),
\end{equation}
and hence
\begin{equation}
		u_I(x,p)
		\longrightarrow
		\Thetaf\!\left(
		\pi I\rhozero(x;I)-|p|
		\right).
	\label{eq:wignerdroplet}
\end{equation}
Note that \eqref{eq:wignerdroplet} should be understood as the leading bulk semiclassical
symbol. Thus the semiclassical support of the rank-$n$ projector is
\begin{equation}
		\mathcal D_I
		=
		\left\{
		(x,p):
		|p|\leq
		\pi I\rhozero(x;I)
		\right\}.
	\label{eq:D-I}
\end{equation}
At each fixed $x$, the droplet therefore extends between
\begin{equation}
	p_-(x;I)
	=
	-\pi I\rhozero(x;I),
	\qquad
	p_+(x;I)
	=
	+\pi I\rhozero(x;I).
	\label{eq:wigner-boundaries}
\end{equation}
These are exactly the momentum branches obtained independently in \eqref{eq:pzeros}.  We have therefore arrived at an important consistency check, that the variable $p$ that entered \eqref{eq:wigner} merely as the Fourier conjugate of the relative coordinate $\eta$
coincides, in the semiclassical limit, with the variable canonically conjugate to the matrix eigenvalue $x$ obtained from the Jacobi recursion.

The projector gives one further piece of information that was not fixed by canonicality alone. Since $\widehat{\Pi}_n$ projects onto exactly $n$ states,
\begin{equation}
	\Tr\widehat{\Pi}_n=n.
\end{equation}
Combining this identity with the Weyl trace formula \eqref{eq:wigner-trace}, and using the fact that $u_I$ approaches unity inside $\mathcal D_I$ and zero outside, gives
\begin{equation}
	n
	=
	\frac{1}{2\pi g_s}
	\int_{\mathcal D_I}\dd x\,\dd p
	=
	\frac{
		\operatorname{Area}(\mathcal D_I)
	}{
		2\pi g_s
	}.
\end{equation}
Therefore,
\begin{equation}
	\operatorname{Area}(\mathcal D_I)
	=
	2\pi n g_s.
\end{equation}
Since $I=ng_s$, we obtain
\begin{equation}
		\operatorname{Area}(\mathcal D_I)
		=
		2\pi I.
	\label{eq:areaI}
\end{equation}
This result has the usual semiclassical interpretation that one quantum state occupies a phase space cell of area $2\pi g_s$, and the rank-$n$ projector fills $n$ such cells. For completeness, the area formula may also be checked directly from the momentum profile. The vertical width of the droplet at fixed $x$ is $p_+(x;I)-p_-(x;I)$, and hence
\begin{align}
	\operatorname{Area}(\mathcal D_I)
	&=
	\int\dd x\,
	\left[
	p_+(x;I)-p_-(x;I)
	\right]
	\nonumber\\
	&=
	2\pi I
	\int\dd x\,\rhozero(x;I)
	\nonumber\\
	&=
	2\pi I,
	\label{eq:area-zero-density}
\end{align}
where in the final step we used the normalization of the limiting zero density. The trace of the projector and the zero-density description thus give the same phase space area.

Finally, for a one-dimensional canonical system the action associated with a closed phase space contour is its enclosed area divided by $2\pi$. Choosing the orientation of $\partial\mathcal D_I$ such that the contour integral is positive, \eqref{eq:areaI} gives
\begin{equation}
		I
		=
		\frac{1}{2\pi}
		\oint_{\partial\mathcal D_I}
		p\,\dd x .
	\label{eq:actionarea}
\end{equation}
This completes a point left open when $I=ng_s$ was first introduced. The recursion algebra established that $I$ is canonically conjugate to the Fourier angle $\theta$ and the projector now shows that its normalization is precisely the standard normalization of an action variable.

The one-cut construction therefore closes into a consistent picture. The Jacobi recursion supplies the canonical pair $(I,\theta)$ and the map to the eigenvalue coordinate $x$. The limiting zero distribution fixes the conjugate momentum through $p_\pm=\pm\pi I\rhozero$.  Finally, the Wigner transform of the Christoffel-Darboux projector independently reproduces the same momentum boundary and shows that the enclosed phase space area is $2\pi I$. The recursion, the polynomial zeros, and the fermionic projector are thus three descriptions of the same semiclassical phase space geometry.

\subsection{Gaussian model}
\label{sec:gaussian}

The Gaussian matrix model provides the simplest explicit realization of the general construction. We take
\begin{equation}
	W(x)=\frac{x^2}{2}.
	\label{eq:gaussian-potential}
\end{equation}
The corresponding orthogonal polynomials are rescaled Hermite polynomials. In the large $N$ scaling used above, their recurrence coefficients are \cite{Marino:2005}
\begin{equation}
	R(\xi)=t\xi,
	\qquad
	s(\xi)=0.
	\label{eq:gaussian-recursion}
\end{equation}
Since $I=t\xi$, the locally frozen Jacobi coefficients defined in \eqref{eq:abdef} become simply
\begin{equation}
	a(I)=\sqrt{I},
	\qquad
	b(I)=0.
\end{equation}
The Jacobi symbol \eqref{eq:jacobisymbol} therefore reduces to
\begin{equation}
		x=2\sqrt{I}\cos\theta .
	\label{eq:gaussian-symbol}
\end{equation}
At fixed $I$, the local Jacobi spectrum is thus the interval
\begin{equation}
	-2\sqrt{I}\leq x\leq2\sqrt{I},
\end{equation}
and the corresponding local density is
\begin{equation}
	\sigma_I(x)
	=
	\frac{
		\Thetaf(4I-x^2)
	}{
		\pi\sqrt{4I-x^2}
	}.
	\label{eq:gaussian-local-density}
\end{equation}

The limiting zero density follows immediately from \eqref{eq:zerodensity}. For a fixed $x$, only recursion levels $I'\geq x^2/4$ contribute, so that
\begin{align}
	\rhozero(x;I)
	&=
	\frac{1}{I}
	\int_0^I\dd I'\,\sigma_{I'}(x)=
	\frac{1}{2\pi I}
	\sqrt{4I-x^2},
	\qquad |x|\leq2\sqrt{I}.
	\label{eq:gaussian-zero-density}
\end{align}
Using the general zero-density relation \eqref{eq:pzeros}, the two momentum branches are therefore
\begin{equation}
		p_\pm(x;I)
		=
		\pm\frac{1}{2}\sqrt{4I-x^2}.
	\label{eq:gaussianp}
\end{equation}
Eliminating $x$ between \eqref{eq:gaussian-symbol} and \eqref{eq:gaussianp} gives
\begin{equation}
		x^2+4p^2=4I .
	\label{eq:ellipse}
\end{equation}
Thus every fixed value of the recursion action $I$ defines an ellipse in the $(x,p)$ plane.  Its semiaxes are $2\sqrt{I}$ and $\sqrt{I}$, and hence its enclosed area is
\begin{equation}
	\operatorname{Area}(\mathcal D_I)
	=
	\pi(2\sqrt{I})(\sqrt{I})
	=
	2\pi I,
\end{equation}
in agreement with the general action relation \eqref{eq:actionarea}.

At the outermost recursion level, $n=N$ and hence $I=t$. The limiting zero density \eqref{eq:gaussian-zero-density} becomes the planar matrix model eigenvalue density,
\begin{equation}
		\rho(x)
		=
		\frac{1}{2\pi t}
		\sqrt{4t-x^2},
		\qquad
		|x|\leq2\sqrt{t},
	\label{eq:gaussian-semicircle}
\end{equation}
which is Wigner's semicircle law in the present normalization \cite{Marino:2005}. Correspondingly, the outer phase space boundary is
\begin{equation}
	p_\pm(x;t)
	=
	\pm\frac{1}{2}\sqrt{4t-x^2}
	=
	\pm\pi t\,\rho(x),
\end{equation}
providing an explicit check of \eqref{eq:pF}. The Gaussian spectral curve,
\begin{equation}
	y^2(z)=z^2-4t,
\end{equation}
gives the same result on the physical cut through $p_\pm=-\ii y_\pm/2$, as required by \eqref{eq:p-y}.

The Gaussian model also makes the fermionic interpretation especially transparent. After the appropriate rescaling of $x$, the functions $\psi_n(x)$ introduced in \eqref{eq:wavefunctions} are the standard harmonic oscillator wavefunctions, as is familiar from the orthogonal polynomial treatment of the Gaussian matrix model
\cite{Ginsparg:1993is}. Consequently, $\widehat{\Pi}_n$ is the projector onto the first $n$ oscillator levels. Applying the Wigner construction of section \ref{sec:wigner-main} to this projector then gives, in the semiclassical limit, the filled elliptical region bounded by \eqref{eq:ellipse}.
\begin{figure}[t]
	\centering
	\includegraphics[width=.82\textwidth]{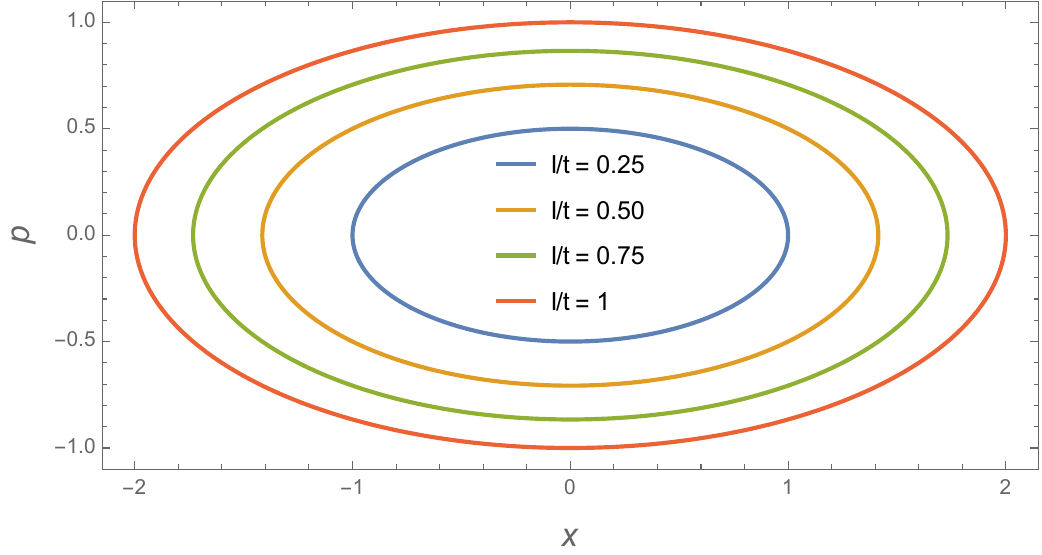}
	\caption{Constant-action curves of the Gaussian matrix model for
		several values of $I/t$, shown in units with $t=1$.  Each curve
		satisfies $x^2+4p^2=4I$ and encloses phase space area $2\pi I$.
		The outermost curve, $I=t$, is the planar phase space boundary
		whose projection onto the $x$ axis has the Wigner semicircle
		density.}
	\label{fig:gaussian}
\end{figure}

\section{The two-cut phase}
\label{sec:twocut}

The one-cut analysis produced a single closed phase space droplet whose outer boundary at $I=t$ is determined by the planar spectral curve and whose enclosed action is $t$. When the planar eigenvalue support splits into disconnected components, one should therefore ask how this picture reorganizes itself. The total 't Hooft coupling is divided among the cuts through the filling fractions, and these filling fractions are themselves period integrals of $y(z)$ or the spectral differential one form $y(z)\, dz$. Since in the one-cut problem we found that the spectral differential is directly related to the canonical momentum, these periods naturally acquire the interpretation of phase space actions. We first make this relation explicit and then show how precisely the same structure emerges from the orthogonal polynomial recursion in the simplest two-cut example.

\subsection{Filling fractions as phase space actions}

Consider a planar solution whose eigenvalue support consists of $s$ disconnected cuts,
\begin{equation}
	\mathcal C
	=
	\bigcup_{a=1}^{s}\mathcal C_a .
\end{equation}
The fraction of eigenvalues occupying the $a$-th cut is
\begin{equation}
	\nu_a
	=
	\int_{\mathcal C_a}\rho(x)\,\dd x,
	\qquad
	\sum_{a=1}^{s}\nu_a=1,
	\label{eq:fillings}
\end{equation}
and it is conventional to introduce the corresponding partial 't Hooft couplings
\begin{equation}
	t_a=t\nu_a,
	\qquad
	\sum_{a=1}^{s}t_a=t.
	\label{eq:partial-tHooft}
\end{equation}
In the spectral-curve description these quantities are the $A$-periods of the planar spectral differential \cite{Dijkgraaf:2002,Eynard:2015},
\begin{equation}
	t_a
	=
	\frac{1}{4\pi\ii}
	\oint_{A_a}y(z)\,\dd z ,
	\label{eq:Aperiod}
\end{equation}
where the cycle $A_a$ surrounds the $a$th cut. The phase space interpretation of this standard relation follows directly from the one-cut result. On a physical cut the two boundary values of the spectral function satisfy
\begin{equation}
	p_\pm(x;t)
	=
	-\frac{\ii}{2}y_\pm(x)
	=
	\pm\pi t\,\rho(x).
	\label{eq:multicut-py}
\end{equation}
Therefore the spectral differential and the canonical one-form are related, on the two sheets surrounding the cut, by
\begin{equation}
	p\,\dd x
	=
	-\frac{\ii}{2}y\,\dd x.
\end{equation}
Substituting this relation into \eqref{eq:Aperiod} gives
\begin{equation}
		t_a
		=
		\frac{1}{2\pi}
		\oint_{A_a}p\,\dd x .
	\label{eq:fillingaction}
\end{equation}
Thus the partial 't Hooft coupling associated with each cut already has precisely the normalization of a one-dimensional action variable.

The same statement may be written in terms of the area enclosed by the corresponding real phase space component $\mathcal D_a$. With the orientation chosen so that the action integral is positive,
\begin{equation}
		\operatorname{Area}(\mathcal D_a)
		=
		2\pi t_a .
	\label{eq:filling-area}
\end{equation}
Consequently,
\begin{equation}
	\sum_{a=1}^{s}
	\operatorname{Area}(\mathcal D_a)
	=
	2\pi t,
\end{equation}
as required by the rank $N$ projector discussed in Sec.~\ref{sec:wigner-main}.

This gives a useful preview of the multicut phase space picture, splitting the eigenvalue support among several cuts corresponds to splitting the total phase space area among several disconnected components. The filling fractions determine how the total action
$t$ is distributed among them.

We now derive this structure directly from the orthogonal polynomial recursion in the simplest nontrivial example.

\subsection{Symmetric quartic double well}

Consider the even quartic potential
\begin{equation}
	W(x)
	=
	\frac{g}{4}x^4-\frac{m}{2}x^2,
	\qquad
	g,m>0.
	\label{eq:doublewell}
\end{equation}
For sufficiently large $m$ relative to the 't Hooft coupling, the potential has two well-separated minima and the planar eigenvalue distribution develops two symmetric cuts. The even symmetry implies that the diagonal recurrence coefficient vanishes,
\begin{equation}
	s_n=0.
\end{equation}
The remaining recurrence coefficient $r_n$ obeys the Freud string equation \cite{DiFrancesco:1993,Bleher:1999},
\begin{equation}
	r_n
	\left[
	-m
	+
	g(r_{n-1}+r_n+r_{n+1})
	\right]
	=
	ng_s .
	\label{eq:freud}
\end{equation}
The important difference from the one-cut phase is that $r_n$ no longer approaches a single smooth function of $n/N$. Instead, in the two-cut regime the leading large $N$ solution alternates between two values. Locally in the slowly varying recursion coordinate
\begin{equation}
	I=ng_s,
\end{equation}
we write
\begin{equation}
	r_{2k}\longrightarrow A(I),
	\qquad
	r_{2k+1}\longrightarrow B(I).
	\label{eq:period2}
\end{equation}
Thus the asymptotic Jacobi operator has a two-site unit cell rather than the one-site translation invariance encountered in the one-cut phase.

The functions $A(I)$ and $B(I)$ follow directly from the string equation. At leading order the difference between $2kg_s$ and $(2k+1)g_s$ is $O(g_s)$, so both even and odd equations may be evaluated at the same continuum value $I$.  For an even site, \eqref{eq:freud} gives
\begin{equation}
	A
	\left[
	-m+g(A+2B)
	\right]
	=
	I,
	\label{eq:freud-even}
\end{equation}
whereas for an odd site it gives
\begin{equation}
	B
	\left[
	-m+g(2A+B)
	\right]
	=
	I.
	\label{eq:freud-odd}
\end{equation}
Subtracting the two equations yields
\begin{equation}
	(A-B)
	\left[
	-m+g(A+B)
	\right]
	=
	0.
\end{equation}
The one-cut solution corresponds to $A=B$.  In the genuine two-cycle
phase $A\neq B$, and therefore
\begin{equation}
	A+B=\frac{m}{g}.
\end{equation}
Substituting this back into either \eqref{eq:freud-even} or \eqref{eq:freud-odd} gives
\begin{equation}
	AB=\frac{I}{g}.
\end{equation}
Thus
\begin{equation}
		A+B=\frac{m}{g},
		\qquad
		AB=\frac{I}{g}.
	\label{eq:AB}
\end{equation}
The two roots are
\begin{equation}
		A,B
		=
		\frac{
			m\pm\sqrt{m^2-4gI}
		}{
			2g
		}.
	\label{eq:ABsol}
\end{equation}
A genuine period-two solution requires the two roots to remain real and distinct,
\begin{equation}
	m^2>4gI.
\end{equation}
Since the largest recursion level is $I=t$, the entire recursion interval remains in the two-cut phase provided
\begin{equation}
		m>2\sqrt{gt}.
	\label{eq:critical}
\end{equation}
At equality the two-cycle degenerates and the two cuts meet at the critical point.

\subsection{Bloch reduction and canonical variables}
\label{sec:bloch-main}

The period-two structure changes the natural Fourier description of
the recursion. In the one-cut phase, freezing the recurrence
coefficients produced a translation-invariant one-site Jacobi operator,
and the Fourier angle $\theta$ diagonalized translation by one lattice
site. In the two-cut phase, however, translation by a single site
exchanges the two members of the unit cell and therefore does not leave
the frozen recurrence relation invariant. The relevant symmetry is
instead translation by an entire two-site cell.

Let $k$ denote the cell index. Since one cell contains two original
recursion sites,
\begin{equation}
	n\simeq 2k .
\end{equation}
The continuum action associated with the cell coordinate is therefore
\begin{equation}
		I_{\rm c}
		\equiv
		g_s k
		=
		\frac{I}{2}+O(g_s),
	\label{eq:cellaction}
\end{equation}
where $I=ng_s$ is the continuum variable associated with the original
recursion level. We will refer to $I_{\rm c}$ as the \emph{cell
	action}. Translation by one cell is diagonalized by a Bloch phase $q$. The same shift-algebra argument used in the one-cut problem then gives
\begin{equation}
		\{q,I_{\rm c}\}=1,
		\qquad
		\Omega=\dd q\wedge\dd I_{\rm c}.
	\label{eq:qIcell}
\end{equation}
Thus $q$ is the angular variable conjugate to the two-site cell coordinate. To see how the two spectral bands arise, we group the two polynomial components in each cell into a two-component Bloch wave. After freezing $A$ and $B$ at fixed $I$, translation by one cell acts as $e^{\ii q}$. The Jacobi recurrence then reduces to the $2\times2$ Bloch symbol
\begin{equation}
	\mathcal J(q)
	=
	\begin{pmatrix}
		0&
		\sqrt A+\sqrt B\,e^{-\ii q}
		\\[1mm]
		\sqrt A+\sqrt B\,e^{\ii q}&
		0
	\end{pmatrix}.
	\label{eq:bloch}
\end{equation}
The detailed construction, including the normalization associated with
the two-site cell, is reviewed in appendix \ref{app:bloch}. The local spectral values are the eigenvalues of $\mathcal J(q)$. They satisfy
\begin{align}
	x^2
	&=
	\left(
	\sqrt A+\sqrt B\,e^{-\ii q}
	\right)
	\left(
	\sqrt A+\sqrt B\,e^{\ii q}
	\right)
	\nonumber\\
	&=
	A+B+2\sqrt{AB}\cos q .
\end{align}
Using \eqref{eq:AB}, this becomes
\begin{equation}
		x_\pm(I,q)
		=
		\pm
		\sqrt{
			\frac{m}{g}
			+
			2\sqrt{\frac{I}{g}}\cos q
		}.
	\label{eq:bands}
\end{equation}
The two signs describe the two Bloch bands. Since $-1\leq\cos q\leq1$, the positive band runs from $a(I)$ to $b(I)$, while the negative band runs from $-b(I)$ to $-a(I)$ so that
\begin{equation}
	\mathcal B_I
	=
	[-b(I),-a(I)]
	\cup
	[a(I),b(I)],
	\label{eq:two-local-bands}
\end{equation}
where the endpoints are
\begin{equation}
		a^2(I)
		=
		\frac{m-2\sqrt{gI}}{g},
		\qquad
		b^2(I)
		=
		\frac{m+2\sqrt{gI}}{g}.
	\label{eq:endpoints}
\end{equation}
Thus the period-two recursion has replaced the single local Jacobi band of the one-cut phase by two disconnected bands. As $I$ increases, the inner endpoint $a(I)$ moves toward the origin, while the outer endpoint $b(I)$ moves outward. At the outermost recursion level $I=t$, these become precisely the endpoints of the planar two-cut solution.

The local density of states follows directly from the Bloch dispersion relation. From
\begin{equation}
	x^2
	=
	\frac{m}{g}
	+
	2\sqrt{\frac{I}{g}}\cos q ,
\end{equation}
we obtain
\begin{equation}
	2x\,\frac{\partial x}{\partial q}
	=
	-2\sqrt{\frac{I}{g}}\sin q .
\end{equation}
Furthermore,
\begin{equation}
	\sin^2 q
	=
	1-
	\frac{
		\left(x^2-m/g\right)^2
	}{
		4I/g
	}.
\end{equation}
Combining these relations gives
\begin{equation}
	\left|
	\frac{\partial q}{\partial x}
	\right|
	=
	\frac{
		2|x|
	}{
		\sqrt{
			4I/g-\left(x^2-m/g\right)^2
		}
	}.
	\label{eq:dqdx}
\end{equation}
Because one Bloch cell contains two original Jacobi sites, the density of states normalized per original site carries an additional factor of one half. The resulting normalized local density is therefore
\begin{equation}
		\sigma_I(x)
		=
		\frac{|x|}
		{\pi
			\sqrt{
				4I/g-\left(x^2-m/g\right)^2
		}},
		\qquad
		a(I)<|x|<b(I).
	\label{eq:period2DOS}
\end{equation}
It is normalized over the two bands together,
\begin{equation}
	\int_{\mathcal B_I}\sigma_I(x)\,\dd x=1.
\end{equation}
Equation~\eqref{eq:dqdx} may therefore be written, after choosing the
two local orientations of the Bloch angle, as
\begin{equation}
	\left.
	\frac{\partial q_\pm}{\partial x}
	\right|_{I_{\rm c}}
	=
	\pm2\pi\sigma_I(x).
	\label{eq:qxdos}
\end{equation}
This is the period-two analogue of the relation between the one-cut
angle and the local arcsine density. It also provides the bridge back
to the canonical phase space construction.

For slowly varying asymptotically periodic recurrence coefficients, the
limiting zero distribution is obtained by averaging the local periodic
density of states along the slow recursion coordinate \cite{VanAssche:1999Periodic}. Thus
\begin{equation}
		\rhozero(x;I)
		=
		\frac{1}{I}
		\int_0^I\dd I'\,
		\sigma_{I'}(x).
	\label{eq:periodic-zero-density}
\end{equation}
Exactly as in the one-cut problem, the limiting zero distribution
selects the momentum
\begin{equation}
		p_\pm(x;I)
		=
		\pm\pi I\,\rhozero(x;I)
		=
		\pm\pi
		\int_0^I\dd I'\,
		\sigma_{I'}(x).
	\label{eq:periodiczero}
\end{equation}
Differentiating at fixed $x$ gives
\begin{equation}
	\left.
	\frac{\partial p_\pm}{\partial I}
	\right|_x
	=
	\pm\pi\sigma_I(x).
\end{equation}
The Bloch angle, however, is conjugate to the cell action
$I_{\rm c}=I/2$. Since $I=2I_{\rm c}$,
\begin{equation}
	\left.
	\frac{\partial p_\pm}{\partial I_{\rm c}}
	\right|_x
	=
	2
	\left.
	\frac{\partial p_\pm}{\partial I}
	\right|_x
	=
	\pm2\pi\sigma_I(x).
\end{equation}
Comparing this result with Eq.~\eqref{eq:qxdos}, we find
\begin{equation}
	\left.
	\frac{\partial p_\pm}{\partial I_{\rm c}}
	\right|_x
	=
	\left.
	\frac{\partial q_\pm}{\partial x}
	\right|_{I_{\rm c}} .
	\label{eq:period2-integrability}
\end{equation}
This is precisely the integrability condition for a local mixed
generating function $S(x,I_{\rm c})$ satisfying
\begin{equation}
	p=\frac{\partial S}{\partial x},
	\qquad
	q=\frac{\partial S}{\partial I_{\rm c}}.
\end{equation}
Consequently, the transformation from the Bloch variables
$(q,I_{\rm c})$ to $(x,p)$ is canonical:
\begin{equation}
		\dd x\wedge\dd p
		=
		\dd q\wedge\dd I_{\rm c},
		\qquad
		\{x,p\}_{q,I_{\rm c}}=1.
	\label{eq:period2canonical}
\end{equation}
The appearance of $I_{\rm c}=I/2$ therefore has a direct geometric
origin. The Bloch phase measures translation between two-site cells, so
its conjugate variable is naturally associated with the cell
coordinate rather than with an individual recursion site.

\subsection{Two droplets and the spectral curve}

We can now evaluate the zero density and the corresponding phase space momentum explicitly. For a fixed value of $x$, the local density \eqref{eq:period2DOS} contributes only when $x$ belongs to one of the two local spectral bands. The support condition is
\begin{equation}
	\frac{4I'}{g}
	\geq
	\left(
	x^2-\frac{m}{g}
	\right)^2 .
\end{equation}
Hence the first recursion level that contributes to the integral is
\begin{equation}
	I_{\min}(x)
	=
	\frac{g}{4}
	\left(
	x^2-\frac{m}{g}
	\right)^2 .
	\label{eq:Imin-two}
\end{equation}
Using \eqref{eq:periodic-zero-density}, we obtain
\begin{align}
	\rhozero(x;I)
	&=
	\frac{|x|}{\pi I}
	\int_{I_{\min}(x)}^{I}
	\frac{\dd I'}
	{
		\sqrt{
			4I'/g-
			\left(x^2-m/g\right)^2
		}
	}=
	\frac{g|x|}{2\pi I}
	\sqrt{
		\frac{4I}{g}
		-
		\left(
		x^2-\frac{m}{g}
		\right)^2
	}.
	\label{eq:twozero-intermediate}
\end{align}
The expression under the square root factorizes in terms of the local band endpoints,
\begin{equation}
	\frac{4I}{g}
	-
	\left(
	x^2-\frac{m}{g}
	\right)^2
	=
	\left[b^2(I)-x^2\right]
	\left[x^2-a^2(I)\right].
\end{equation}
Therefore
\begin{equation}
		\rhozero(x;I)
		=
		\frac{g|x|}{2\pi I}
		\sqrt{
			\left[b^2(I)-x^2\right]
			\left[x^2-a^2(I)\right]
		}.
	\label{eq:twozero}
\end{equation}
The density has support on the two intervals $[-b(I),-a(I)]\cup[a(I),b(I)]$ and vanishes at both the inner and outer endpoints.

The corresponding phase space momentum follows immediately from \eqref{eq:periodiczero},
\begin{equation}
		p_\pm(x;I)
		=
		\pm\frac{g|x|}{2}
		\sqrt{
			\left[b^2(I)-x^2\right]
			\left[x^2-a^2(I)\right]
		}.
	\label{eq:twop}
\end{equation}
Thus the allowed region in the $(x,p)$ plane is
\begin{equation}
	|p|
	\leq
	\frac{g|x|}{2}
	\sqrt{
		\left[b^2(I)-x^2\right]
		\left[x^2-a^2(I)\right]
	},
\end{equation}
with $a(I)<|x|<b(I)$. Since the interval around the origin is excluded, this phase space region has two disconnected components. One lying over the positive-$x$ band and one over the negative-$x$ band. The single droplet of the one-cut phase has therefore split into two disconnected droplets.

At the outermost recursion level $I=t$, \eqref{eq:twozero} becomes the planar matrix-model density,
\begin{equation}
		\rho(x)
		=
		\frac{g|x|}{2\pi t}
		\sqrt{
			(b^2-x^2)(x^2-a^2)
		}.
	\label{eq:twodensity}
\end{equation}
Here $a=a(t)$ and $b=b(t)$. The corresponding planar spectral curve is
\begin{equation}
		y^2(z)
		=
		g^2z^2
		(z^2-a^2)(z^2-b^2).
	\label{eq:twocurve}
\end{equation}
On either physical cut its boundary values satisfy
\begin{equation}
	p_\pm(x;t)
	=
	-\frac{\ii}{2}y_\pm(x)
	=
	\pm\pi t\,\rho(x),
\end{equation}
which reproduces \eqref{eq:twop} at $I=t$. Thus the identification between the real phase space boundary and the boundary values of the spectral curve found in the one-cut problem survives unchanged. What changes in the two-cut phase is the topology of the real support, there are now two disconnected components. It remains to determine the action carried by each component. Since the potential and the two-cut solution are symmetric under $x\rightarrow-x$, the limiting zero density is even,
\begin{equation}
	\rhozero(-x;I)=\rhozero(x;I).
\end{equation}
Together with its normalization, this implies
\begin{equation}
	\int_{a(I)}^{b(I)}
	\rhozero(x;I)\,\dd x
	=
	\int_{-b(I)}^{-a(I)}
	\rhozero(x;I)\,\dd x
	=
	\frac12 .
\end{equation}
The phase space area of the positive component is therefore
\begin{align}
	\operatorname{Area}(\mathcal D_+)
	&=
	\int_{a(I)}^{b(I)}
	\dd x\,
	\left[
	p_+(x;I)-p_-(x;I)
	\right]
	\nonumber\\
	&=
	2\pi I
	\int_{a(I)}^{b(I)}
	\dd x\,\rhozero(x;I)
	\nonumber\\
	&=
	\pi I.
\end{align}
By symmetry the negative component has the same area,
\begin{equation}
		\operatorname{Area}(\mathcal D_+)
		=
		\operatorname{Area}(\mathcal D_-)
		=
		\pi I.
	\label{eq:two-component-areas}
\end{equation}
Since $I_{\rm c}=I/2$, each component therefore encloses
\begin{equation}
	\operatorname{Area}(\mathcal D_\pm)
	=
	2\pi I_{\rm c}.
\end{equation}
To distinguish the geometrical actions of the two disconnected components from the cell action introduced above, let us define
\begin{equation}
	I_\pm
	\equiv
	\frac{1}{2\pi}
	\oint_{\partial\mathcal D_\pm}
	p\,\dd x ,
	\label{eq:droplet-actions-def}
\end{equation}
with the contour orientations chosen so that the integrals are positive. Their areas then give
\begin{equation}
		I_+
		=
		I_-
		=
		\frac{I}{2}
		=
		I_{\rm c}.
	\label{eq:twoactions}
\end{equation}
This gives a direct geometric interpretation of the cell action that appeared in the Bloch construction. In the symmetric period-two problem, the variable canonically conjugate to the Bloch angle is precisely the action enclosed by either of the two disconnected phase space components. At the outermost recursion level $I=t$,
\begin{equation}
	I_+=I_-=\frac{t}{2}.
\end{equation}
The symmetric two-cut solution also has equal filling fractions,
\begin{equation}
	\nu_+=\nu_-=\frac12,
\end{equation}
and therefore
\begin{equation}
	t_+=t_-=\frac{t}{2}.
\end{equation}
The phase space actions of the two outermost droplets therefore coincide with the corresponding partial 't Hooft couplings
\begin{equation}
		I_\pm
		=
		t_\pm
		=
		\frac{1}{2\pi}
		\oint_{A_\pm}p\,\dd x
		=
		\frac{t}{2}.
	\label{eq:two-filling-action}
\end{equation}
Thus the orthogonal polynomial recursion, the Bloch reduction, the limiting zero distribution, and the spectral-curve filling fractions all describe the same decomposition of the total action.

The two-cut phase therefore provides the first nontrivial extension of the one-cut picture. The period-two Jacobi recursion introduces a two-site unit cell and replaces the one-site Fourier angle $\theta$ by a Bloch angle $q$ conjugate to the cell action $I_{\rm c}=I/2$. The two eigenvalues of the Bloch-reduced Jacobi symbol produce two local spectral bands. Their averaged density of states determines the zero distribution and, through
$p_\pm=\pm\pi I\rhozero$, two disconnected phase space droplets. In the symmetric quartic model each droplet carries the action
\begin{equation}
	I_\pm=I_{\rm c}=\frac{I}{2}.
\end{equation}
At the outermost level $I=t$, these actions become the partial 't Hooft couplings,
\begin{equation}
	I_\pm=t_\pm=\frac{t}{2}.
\end{equation}
The splitting of the planar eigenvalue support into two cuts is thus
accompanied by a corresponding splitting of the total phase space
action into two equal components.

\begin{figure}[t]
	\centering
	\includegraphics[width=.82\textwidth]{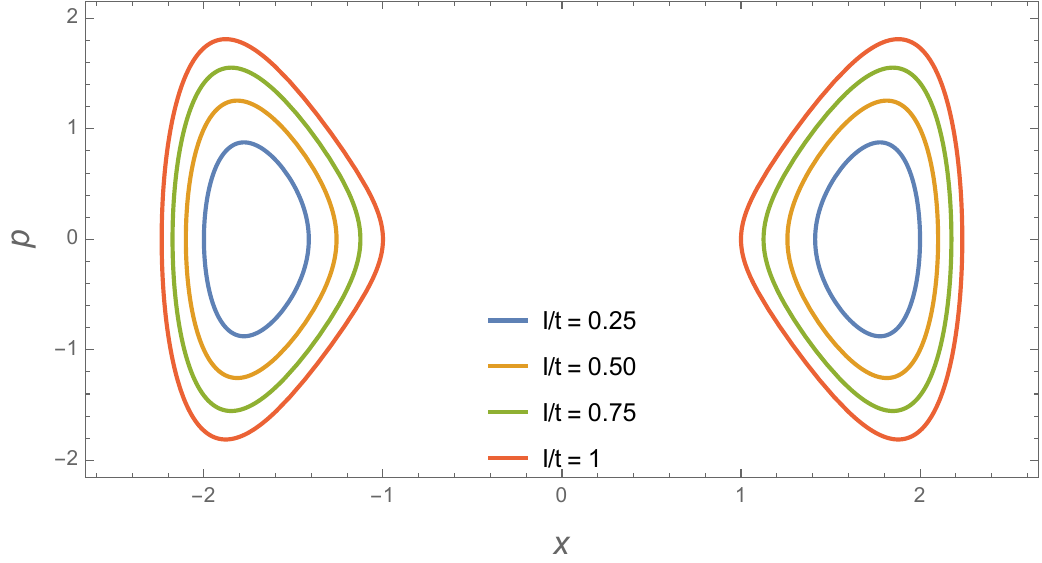}
	\caption{
		Constant-$I$ phase space curves of the symmetric quartic
		two-cut model for $g=1$, $m=3$ and $t=1$. For every
		$I\leq t$ the allowed region consists of two disconnected
		components associated with the positive and negative local
		Jacobi bands. Increasing $I$ moves the inner endpoints toward
		the origin and the outer endpoints outward. Each component
		encloses area $\pi I=2\pi I_{\rm c}$, where
		$I_{\rm c}=I/2$ is the cell action conjugate to the Bloch
		angle $q$. At $I=t$ the two components carry the equal
		filling-fraction actions $t_+=t_-=t/2$.
	}
	\label{fig:twocut}
\end{figure}

\section{Multicut phases and finite-gap recursion}
\label{sec:multicut}

For a general planar solution with $s$ disconnected real cuts, let
\begin{equation}
	\mathcal C
	=
	\bigcup_{a=1}^{s}
	[x_{2a},x_{2a-1}] .
	\label{eq:multicut-support}
\end{equation}
Assuming the branch points are simple and the solution is away from critical degenerations, the planar spectral curve takes the hyperelliptic form
\begin{equation}
		y^2(z)
		=
		M^2(z)
		\prod_{\alpha=1}^{2s}
		(z-x_\alpha).
	\label{eq:multicurve}
\end{equation}
The $2s$ branch points determine the endpoints of the $s$ cuts. The essential phase space interpretation developed in the one and two-cut examples continues to hold cut-by-cut. On every physical cut, the two boundary values of the spectral function satisfy
\begin{equation}
		p_\pm(x)
		=
		-\frac{\ii}{2}y_\pm(x)
		=
		\pm\pi t\,\rho(x).
	\label{eq:multipp}
\end{equation}
We may therefore associate with each connected component of the eigenvalue support a corresponding real phase space component bounded by these two branches. Let
\begin{equation}
	\nu_a
	=
	\int_{\mathcal C_a}\rho(x)\,\dd x,
	\qquad
	t_a=t\nu_a,
\end{equation}
denote the filling fraction and partial 't Hooft coupling associated with the $a$-th cut.  Since the vertical width of the corresponding phase space component is
\begin{equation}
	p_+(x)-p_-(x)
	=
	2\pi t\,\rho(x),
\end{equation}
its area is
\begin{align}
	\operatorname{Area}(\mathcal D_a)
	&=
	\int_{\mathcal C_a}
	\dd x\,
	[p_+(x)-p_-(x)]
	\nonumber\\
	&=
	2\pi t
	\int_{\mathcal C_a}\rho(x)\,\dd x
	\nonumber\\
	&=
2\pi t_a.
	\label{eq:multiarea}
\end{align}
Equivalently,
\begin{equation}
		t_a
		=
		\frac{1}{2\pi}
		\oint_{\partial\mathcal D_a}
		p\,\dd x ,
	\label{eq:multicut-action}
\end{equation}
with the contour oriented so that the integral is positive. The partial 't Hooft couplings are therefore precisely the actions carried by the individual phase space components. Their sum reproduces the total action,
\begin{equation}
	\sum_{a=1}^{s}t_a=t,
	\qquad
	\sum_{a=1}^{s}
	\operatorname{Area}(\mathcal D_a)
	=
	2\pi t.
\end{equation}
Once the total coupling $t$ is fixed, only $s-1$ filling fractions are independent. This is also the genus of the hyperelliptic curve \eqref{eq:multicurve},
\begin{equation}
	g_{\rm curve}=s-1.
\end{equation}
This agreement counts the independent period data that determine how the total action is distributed among the cuts. It should not, however, be interpreted as introducing $s-1$ additional canonical momenta on each real phase space component. Each connected droplet remains a two-dimensional phase space region with its own action. The additional data specify the relative distribution of the total action among the different components.

The orthogonal polynomial description of the same multicut geometry is less elementary.  In the one-cut phase, the recurrence coefficients approached a single slowly varying pair $R(\xi)$ and $s(\xi)$. In the symmetric two-cut example, this was replaced by a slowly varying period-two pattern, which could be treated by an ordinary two-component Bloch analysis. For more general multicut solutions, however, neither description is generic.

A useful intermediate class is provided by asymptotically periodic recurrence coefficients. Suppose that, after freezing the slow recursion coordinate $I$, the Jacobi coefficients have period $\mr$. We then group $\mr$ consecutive recursion sites into a unit cell.

To see how the corresponding spectral bands arise, it is useful to write the frozen Jacobi equation in transfer-matrix form. Introducing
\begin{equation}
	a_n=\sqrt{r_n},
	\qquad
	b_n=-s_n,
\end{equation}
the recurrence relation
\begin{equation}
	a_{n+1}\psi_{n+1}
	+b_n\psi_n
	+a_n\psi_{n-1}
	=
	x\psi_n
\end{equation}
may be written as
\begin{equation}
	\begin{pmatrix}
		\psi_{n+1}\\
		\psi_n
	\end{pmatrix}
	=
	T_n(x)
	\begin{pmatrix}
		\psi_n\\
		\psi_{n-1}
	\end{pmatrix},
\end{equation}
where
\begin{equation}
	T_n(x)
	=
	\begin{pmatrix}
		\dfrac{x-b_n}{a_{n+1}}
		&
		-\dfrac{a_n}{a_{n+1}}
		\\[3mm]
		1&0
	\end{pmatrix}.
\end{equation}
Propagation through one complete $\mr$-site cell is therefore described
by the monodromy matrix
\begin{equation}
	\mathcal M_\mr(x;I)
	=
	T_{n+\mr-1}(x)\cdots T_n(x).
\end{equation}
Because the coefficients are periodic, $a_{n+\mr}=a_n$, its determinant is unity,
\begin{equation}
	\det\mathcal M_\mr
	=
	\prod_{j=0}^{\mr-1}
	\frac{a_{n+j}}{a_{n+j+1}}
	=
	1.
\end{equation}
Translation by one complete unit cell is diagonalized by a Bloch phase $q_{\rm B}$,
\begin{equation}
	\psi_{n+\mr}
	=
	e^{\ii q_{\rm B}}\psi_n.
\end{equation}
Thus the two eigenvalues of $\mathcal M_\mr$ are $e^{\pm\ii q_{\rm B}}$. Defining the discriminant of the periodic Jacobi problem as the trace of the one-cell transfer matrix,
\begin{equation}
		\Delta_\mr(x;I)
		\equiv
		\Tr\mathcal M_\mr(x;I),
\end{equation}
we obtain
\begin{equation}
		\Delta_\mr(x;I)
		=
		e^{\ii q_{\rm B}}+e^{-\ii q_{\rm B}}
		=
		2\cos q_{\rm B}.
	\label{eq:discriminant}
\end{equation}
The meaning of this relation is particularly simple. For a real Bloch angle the eigenvalues of the transfer matrix have unit modulus, so the solutions oscillate rather than grow or decay from cell to cell. Real Bloch phases therefore exist precisely when
\begin{equation}
	|\Delta_\mr(x;I)|\leq2,
\end{equation}
and these intervals constitute the allowed local spectral bands. The band edges occur at $\Delta_\mr=\pm2$, corresponding to $q_{\rm B}=0$ or $\pi$. The local density of states follows directly by differentiating the Bloch relation. Since
\begin{equation}
	-\sin q_{\rm B}\,
	\frac{\partial q_{\rm B}}{\partial x}
	=
	\frac12
	\frac{\partial\Delta_r}{\partial x},
\end{equation}
and
\begin{equation}
	|\sin q_{\rm B}|
	=
	\frac12
	\sqrt{4-\Delta_\mr^2},
\end{equation}
we obtain
\begin{equation}
	\left|
	\frac{\partial q_{\rm B}}{\partial x}
	\right|
	=
	\frac{
		|\partial_x\Delta_\mr(x;I)|
	}{
		\sqrt{4-\Delta_\mr^2(x;I)}
	}.
\end{equation}
Because one Bloch cell contains $\mr$ original Jacobi sites, the density normalized per original site is
\begin{equation}
		\sigma_I(x)
		=
		\frac{1}{\mr\pi}
		\frac{
			|\partial_x\Delta_r(x;I)|
		}{
			\sqrt{4-\Delta_\mr^2(x;I)}
		},
		\qquad
		|\Delta_\mr(x;I)|<2 .
	\label{eq:generalDOS}
\end{equation}
The constructions studied above are recovered as the first two cases: $\mr=1$ gives the local arcsine density of the one-cut phase, whereas $\mr=2$ reproduces the period-two density \eqref{eq:period2DOS} of the symmetric quartic model.

When the periodic coefficients also vary slowly with the recursion level, the limiting zero distribution is again obtained by averaging this local density of states over the slow coordinate \cite{VanAssche:1999Periodic}. Thus the mechanism encountered in the one and two-cut examples persists. The local spectral problem determines $\sigma_I(x)$, while its accumulated contribution along the recursion determines the macroscopic distribution of polynomial zeros. 

A generic multicut matrix model, however, need not correspond to a finite-period Jacobi operator. The discrete occupation numbers of the different cuts lead, for generic filling fractions, to oscillatory and typically quasiperiodic large $N$ recurrence coefficients
\cite{Bonnet:2000}. The appropriate mathematical framework is then finite-gap Jacobi theory rather than ordinary finite-cell Bloch theory. The different phases of the quasiperiodic recurrence coefficients live on the associated isospectral torus, and related finite-gap structures also appear in continuum Toda descriptions \cite{Kuijlaars:2000}.

The physical picture is nevertheless already clear from the spectral curve. The $s$ cuts determine $s$ disconnected real phase space droplets, and their actions are the partial 't Hooft couplings $t_a$. What becomes more involved in the generic multicut problem is
not the identification of these actions, but the reconstruction of the full canonical map directly from the orthogonal polynomial recursion. For finite-period recurrences this reconstruction reduces to ordinary Bloch theory. Developing that general canonical construction would require the full machinery of finite-gap Jacobi theory and is kept for a future work.

\section{Summary and outlook}
\label{sec:conclusion}

We have formulated a phase space description of the large $N$ Hermitian one matrix model using structures intrinsic to its orthogonal polynomial solution. In the regular one-cut phase, the polynomial degree defines the continuum variable $I=ng_s$, while the Fourier phase of the Jacobi recursion is its canonical conjugate, $\{\theta,I\}=1$. The Jacobi symbol maps this pair to the matrix eigenvalue coordinate, $x=b(I)+2a(I)\cos\theta$, and a generating function constructs the conjugate momentum while preserving the symplectic form. The resulting momentum has a direct orthogonal polynomial meaning,
\begin{equation}
	p_\pm(x;I)=\pm\pi I\rhozero(x;I).
\end{equation}
Thus the zeros of $p_n$ are not a discrete momentum lattice. Instead, their limiting density, or equivalently their local bulk spacing, reconstructs the continuous momentum on the constant-$I$ curve. The complete triangular array of zeros for $n=1,\ldots,N$ may therefore be viewed as a coordinate-space for the nested phase space levels.

The same boundary follows independently from the Wigner transform of the Christoffel-Darboux projector. In the semiclassical limit the rank-$n$ projector fills a region of area $2\pi I$, which verifies that $I$ has the standard normalization of an action variable. At the outermost level $I=t$, the zero density becomes the planar eigenvalue density and
\begin{equation}
	p_F(x)=\pi t\rho(x)=-\frac{\ii}{2}y_+(x).
\end{equation}
The planar spectral curve therefore provides the complex continuation of the real phase space boundary.

The symmetric quartic double well gives a nontrivial extension. Its two-cut phase is encoded by a slowly varying period-two Jacobi recursion. The associated Bloch problem reproduces the two-cuts and the planar density. Because the Bloch angle is conjugate to a two-site cell, the natural cell action is $I_c=I/2$; the two disconnected components each have area $2\pi I_c$, and at $I=t$ these actions coincide with the partial 't Hooft couplings. More generally,
\begin{equation}
	t_a=\frac1{2\pi}\oint_{A_a}p\,\dd x
\end{equation}
is the standard filling-fraction period written directly as an action integral.

The contrast with the unitary matrix model is also instructive. In a unitary model the eigenvalue coordinate is itself an angle, while the conjugate momentum direction is naturally discrete. In the phase space constructions of
\cite{Chattopadhyay:2017,Chattopadhyay:2018}, Young-diagram data provide
this momentum directly. For a Hermitian model the eigenvalue coordinate instead lies on $\mathbb R$, and the conjugate momentum is continuous. The discrete datum carried by the orthogonal polynomial description is the polynomial degree $n$, which becomes the action variable $I=ng_s$ in the semiclassical limit, while the momentum is reconstructed
from the spatial oscillation and zero distribution of the polynomial states. In this sense the two matrix model classes organize their discrete and continuous phase space data in complementary ways.

The phase space formulation developed here brings together several structures which are usually discussed separately. The Jacobi recursion supplies the action-angle variables, the zero distribution determines the momentum profile, and the Christoffel-Darboux projector
provides an underlying fermionic description whose semiclassical Wigner symbol fills the corresponding phase space region. At the outermost level, the planar spectral curve becomes the boundary of this occupied region, while in the multicut problem its periods become the actions of the individual components. In this sense the action-angle language
provides a direct bridge between the orthogonal polynomial, free fermion, and spectral curve descriptions of the matrix model.

This viewpoint also suggests a natural way of approaching finite $N$ physics. The Jacobi matrix, the polynomial zeros, and the rank $n$ projector are all exact finite dimensional objects, whereas the droplet appears only in the semiclassical scaling
$N,n\rightarrow\infty$, $g_s\rightarrow0$ with $I=ng_s$ fixed. Therefore this work provides a possible starting point for studying the departure from the classical droplet through the Moyal expansion, together with the finite $N$ properties of the recurrence coefficients and the Christoffel-Darboux kernel. It would be interesting to understand how these corrections are related to the usual genus expansion and to topological recursion.  Near the phase space boundary one should expect the ordinary Moyal derivative expansion to become non-uniform, which should also be taken care of for studying a complete finite $N$ description.

A further direction is suggested by the Wigner formulation itself. Instead of the rank $n$ projector, one may study the Wigner functions of the individual orthogonal polynomial states $\psi_n$, or finite superpositions of them, and quantify their departure from a classical positive phase space distribution. In the Gaussian model these states reduce to harmonic oscillator levels, providing an immediate setting in which Wigner negativity can be studied. Whether such quantities admit a useful interpretation in terms of different notions of nonclassicality, including Wigner negativity or ``magic'' in continuous-variable quantum information, would be interesting to explore. Other natural extensions include the full canonical treatment of quasiperiodic multicut recursions and the interpretation of cut-merging critical points as singular limits of the phase space boundary.

\section*{Acknowledgement} The author thanks Suvankar Dutta (IISER Bhopal) and Parikshit Dutta (Asutosh College) for several useful discussions and introducing the author to the wonderful world of matrix models. The author is also grateful to Debangshu Mukherjee (POSTECH) for careful reading of the manuscript and for several helpful comments and corrections on the draft. OpenAI’s ChatGPT was used to assist with preparation of notes as well as language refinement during the preparation of this work. This work is supported by U.S.A. National Science Foundation Award OAC-$2334265$.

\appendix

\section{Christoffel-Darboux kernel}
\label{app:cd}

Algebraically, the Christoffel-Darboux is the integral kernel of the projection onto the first $n$ orthogonal polynomial states \eqref{eq:kernel}. In the large $n$ limit its short-distance form becomes universal, and that universal sine kernel is precisely what turns the Wigner transform of the projector into a momentum interval.

\subsection{From the three-term recurrence to the Christoffel-Darboux identity}

For the orthonormal polynomials of section \ref{sec:op-prelim}, define the unweighted kernel
\begin{equation}
	\mathcal K_n(x,y)=\sum_{k=0}^{n-1}P_k(x)P_k(y).
	\label{eq:app-unweighted-kernel}
\end{equation}
Using the recurrence \eqref{eq:recurrence} once at $x$ and once at $y$, one finds
\begin{align}
	(x-y)P_k(x)P_k(y)
	&=\sqrt{r_{k+1}}
	\bigl[P_{k+1}(x)P_k(y)-P_k(x)P_{k+1}(y)\bigr]
	\nonumber\\
	&\quad
	-\sqrt{r_k}
	\bigl[P_k(x)P_{k-1}(y)-P_{k-1}(x)P_k(y)\bigr].
	\label{eq:app-cd-telescope-one}
\end{align}
Summing from $k=0$ to $n-1$, leaving only the upper endpoint,
\begin{equation}
		\mathcal K_n(x,y)
		=\sqrt{r_n}\,
		\frac{P_n(x)P_{n-1}(y)-P_{n-1}(x)P_n(y)}{x-y}.
	\label{eq:app-CD}
\end{equation}
This is the Christoffel-Darboux identity \cite{Simon:2008}. The weighted kernel used in the main text is obtained by multiplying by the wavefunction factors,
\begin{equation}
	K_n(x,y)
	=\frac{e^{-[W(x)+W(y)]/(2g_s)}}{2\pi}\,
	\mathcal K_n(x,y).
	\label{eq:app-weighted-CD}
\end{equation}
In particular, the diagonal value $K_n(x,x)$ is the local one-particle density of the rank-$n$ Slater state.

\subsection{Eigenvalue correlations}

For $n=N$, the joint eigenvalue probability density is the squared Slater determinant in \eqref{eq:slater}. Integrating out $N-k$ eigenvalues gives the standard determinantal $k$-point correlation function
\begin{equation}
	R_k(x_1,\ldots,x_k)
	=\det_{1\leq i,j\leq k}K_N(x_i,x_j).
	\label{eq:app-determinantal}
\end{equation}
Thus
\begin{equation}
	R_1(x)=K_N(x,x),
\end{equation}
and the normalized macroscopic eigenvalue density is $K_N(x,x)/N$ in the large $N$ limit. More generally, keeping $n/N=\xi$ fixed, the rank-$n$ kernel defines an $n$-particle orthogonal polynomial ensemble with the same one-body weight $e^{-W/g_s}$. Since
$I=ng_s=t\xi$, its large-$n$ saddle equation is the usual planar equation with $t$ replaced by $I$,
\begin{equation}
	\frac{1}{2I}W'(x)
	=
	\mathcal P\int\frac{\rho_I(x')\,\dd x'}{x-x'}.
	\label{eq:app-partial-saddle}
\end{equation}
The standard orthogonal polynomial analysis of this partial kernel gives
\begin{equation}
	\rho_I(x)
	=\lim\frac{K_n(x,x)}{n}
	=\frac1I\int_0^I\dd I'\,\sigma_{I'}(x).
	\label{eq:app-partial-density}
\end{equation}
Appendix \ref{app:kva} shows that the same averaged local measure is
the limiting zero density $\rhozero(x;I)$. Thus
$\rho_I=\rhozero$ in the regular regime considered here, while at the
outermost level $I=t$ both reduce to the physical planar density
$\rho(x)$.

\subsection{Bulk scaling and the sine kernel}

Let $x$ lie in a regular bulk region where $\rhozero(x;I)>0$ and varies only on a macroscopic scale. The mean spacing of the $n$ points near $x$ is
\begin{equation}
	\Delta x_{\rm micro}\sim\frac1{n\rhozero(x;I)}.
	\label{eq:app-spacing-scale}
\end{equation}
Bulk universality of unitary orthogonal polynomial ensembles states that after distances are measured in units of this local spacing, the properly normalized Christoffel-Darboux kernel tends to the sine kernel at regular bulk points for the class of varying exponential
weights relevant here \cite{Deift:1999Varying,Lubinsky:2016,mclaughlin2008}:
\begin{equation}
		\frac{1}{n\rhozero(x;I)}
		K_n\!\left(
		x+\frac{u}{n\rhozero(x;I)},
		x+\frac{v}{n\rhozero(x;I)}
		\right)
		\longrightarrow
		\frac{\sin\pi(u-v)}{\pi(u-v)}.
	\label{eq:app-sine-scaled}
\end{equation}
The result is universal in the sense that the detailed potential enters at leading bulk order only through the local density used to unfold the spectrum. It is the same sine kernel that appears throughout the bulk of unitary random-matrix ensembles.

For the Wigner transform it is convenient to center the two arguments around the same macroscopic point,
\begin{equation}
	x_1=x+\frac\eta2,
	\qquad
	x_2=x-\frac\eta2.
\end{equation}
Setting $u-v=n\rhozero(x;I)\eta$ in Eq.~\eqref{eq:app-sine-scaled} gives the form used in the main text,
\begin{equation}
	\boxed{
		K_n\!\left(x+\frac\eta2,x-\frac\eta2\right)
		\simeq
		\frac{\sin[\pi n\rhozero(x;I)\eta]}{\pi\eta}.}
	\label{eq:app-sine-eta}
\end{equation}
The use of this expression in the Wigner transform is therefore to be understood in the local semiclassical sense, no uniform extension of the sine-kernel approximation to macroscopic separations is assumed. The diagonal limit $\eta\to0$ is
\begin{equation}
	K_n(x,x)\simeq n\rhozero(x;I),
\end{equation}
consistent with the interpretation above.

A useful physical way to read Eq.~\eqref{eq:app-sine-eta} is that the projector has a local Fermi wave number
\begin{equation}
	k_F(x;I)=\pi n\rhozero(x;I).
	\label{eq:app-kf}
\end{equation}
The kernel is locally the coordinate-space kernel of a one-dimensional filled interval of Fourier modes $|k|\leq k_F$. Since the Wigner transform uses the dimensionful Fourier variable $p=g_sk$, this immediately predicts
\begin{equation}
	p_F(x;I)=g_s k_F(x;I)=\pi I\rhozero(x;I),
\end{equation}
which is exactly the canonical momentum found in the main text. appendix \ref{app:wigner} performs the Fourier transform with all normalization factors included. 

One should note that \eqref{eq:app-sine-scaled} is a bulk statement. Near a regular spectral edge the density vanishes and the microscopic scaling changes from the sine kernel to the Airy kernel and at multicritical points still different kernels appear. For the present purpose this distinction is important but not problematic. The sine kernel determines the interior of the semiclassical phase space region and hence its leading sharp momentum boundary. The detailed smoothing of that boundary is a finite-$g_s$ effect and belongs to the edge scaling problem, which we leave for future work.

\section{Kuijlaars-Van Assche theorem}
\label{app:kva}

This appendix records the precise zero-distribution theorem used in Sec.~\ref{sec:zeros-momentum} and explains its translation to the matrix-model notation. The double-index notation with both $n$ and $N$ is useful here because the orthogonality measure itself varies with $N$. In the main text this dependence is suppressed in accordance with standard matrix model notation \cite{Marino:2005}.

\subsection{Statement of the theorem}

Consider, for every positive integer $N$, a family of orthogonal polynomials $p_{n,N}$ generated by
\begin{equation}
	x p_{n,N}(x)
	=a_{n+1,N}p_{n+1,N}(x)
	+b_{n,N}p_{n,N}(x)
	+a_{n,N}p_{n-1,N}(x),
	\qquad n\geq0,
	\label{eq:kva-rec}
\end{equation}
with $a_{n,N}>0$, $b_{n,N}\in\mathbb R$, $p_{0,N}=1$ and $p_{-1,N}=0$. Let $\nu(p_{n,N})$ denote the normalized zero-counting measure,
\begin{equation}
	\nu(p_{n,N})=\frac1n\sum_{j=1}^{n}\delta_{x_{j,n}^{(N)}}.
	\label{eq:kva-zero-measure}
\end{equation}
Suppose there exist continuous functions
\begin{equation}
	a:(0,\infty)\to[0,\infty),
	\qquad
	b:(0,\infty)\to\mathbb R,
\end{equation}
such that, whenever $n,N\to\infty$ with $n/N\to\tau>0$,
\begin{equation}
	a_{n,N}\longrightarrow a(\tau),
	\qquad
	b_{n,N}\longrightarrow b(\tau).
	\label{eq:kva-limits}
\end{equation}
Define
\begin{equation}
	\alpha(\tau)=b(\tau)-2a(\tau),
	\qquad
	\beta(\tau)=b(\tau)+2a(\tau).
	\label{eq:kva-edges}
\end{equation}
Theorem~1.4 of Kuijlaars and Van Assche \cite{Kuijlaars:1999}
states that the zero-counting measures converge \emph{weakly},
\begin{equation}
		\underset{n/N\to\tau}{\mathrm{w\!-\!lim}}\,
		\nu(p_{n,N})
		=\frac1\tau\int_0^\tau
		\omega_{[\alpha(u),\beta(u)]}\,\dd u.
	\label{eq:kva-theorem}
\end{equation}
Equivalently, for every continuous test function $F$ of compact support,
\begin{equation}
	\lim_{n/N\to\tau}\frac1n\sum_{j=1}^{n}
	F\!\left(x_{j,n}^{(N)}\right)
	=
	\frac1\tau\int_0^\tau\dd u
	\int F(x)\,\dd\omega_{[\alpha(u),\beta(u)]}(x).
	\label{eq:kva-test-function}
\end{equation}
Here $\omega_{[\alpha,\beta]}$ is the arcsine measure on the interval
$[\alpha,\beta]$,
\begin{equation}
	\dd\omega_{[\alpha,\beta]}(x)
	=\frac{\Thetaf((\beta-x)(x-\alpha))}
	{\pi\sqrt{(\beta-x)(x-\alpha)}}\,\dd x,
	\qquad \alpha<\beta,
	\label{eq:kva-arcsine}
\end{equation}
with the degenerate interval interpreted as a Dirac mass. The same
measure can be written as the pushforward of the uniform measure
$\dd\vartheta/\pi$ on $0\leq\vartheta\leq\pi$ under
$x=b+2a\cos\vartheta$. Thus the limiting zero measure is an average of
the equilibrium, equivalently arcsine, measures associated with the continuously varying local recurrence coefficients. 

\subsection{Mapping to the Hermitian matrix model}

The zeros do not depend on the normalization of the polynomial. We may therefore apply \eqref{eq:kva-theorem} to the orthonormal family $P_n$ of the matrix model, whose recurrence in the main text is
\begin{equation}
	xP_n=-s_nP_n+\sqrt{r_{n+1}}P_{n+1}+\sqrt{r_n}P_{n-1}.
\end{equation}
Restoring the suppressed $N$ dependence, the identification with \eqref{eq:kva-rec} is
\begin{equation}
	a_{n,N}^{\rm KVA}=\sqrt{r_{n,N}},
	\qquad
	b_{n,N}^{\rm KVA}=-s_{n,N}.
	\label{eq:kva-map-coeff}
\end{equation}
Indeed, the coefficient multiplying $P_{n+1,N}$ is $a_{n+1,N}^{\rm KVA}=\sqrt{r_{n+1,N}}$, while that multiplying $P_{n-1,N}$ is $a_{n,N}^{\rm KVA}=\sqrt{r_{n,N}}$, exactly matching the indexing of the matrix-model recurrence. The smooth large $N$ limit used in the one-cut phase,
\begin{equation}
	r_{n,N}\to R(\xi),
	\qquad
	s_{n,N}\to s(\xi),
	\qquad
	\xi=\frac nN,
\end{equation}
therefore gives
\begin{equation}
	a(u)=\sqrt{R(u)},
	\qquad
	b(u)=-s(u).
	\label{eq:kva-map-limit}
\end{equation}
The local interval in the theorem is consequently
\begin{equation}
	-s(u)-2\sqrt{R(u)}
	\leq x\leq
	-s(u)+2\sqrt{R(u)},
\end{equation}
and the density of its arcsine measure is
\begin{equation}
	\sigma_u(x)=\frac1\pi
	\frac{\Thetaf\!\left(4R(u)-[x+s(u)]^2\right)}
	{\sqrt{4R(u)-[x+s(u)]^2}}.
	\label{eq:kva-sigma-u}
\end{equation}
This is not a new density. It is exactly the density of states of the Jacobi operator whose coefficients were locally frozen in section \ref{sec:canonical-pair}. To see this explicitly, set $I'=tu$ in \eqref{eq:abdef} to write
\begin{equation}
	a(I')=\sqrt{R(u)},
	\qquad
	b(I')=-s(u),
\end{equation}
and therefore
\begin{equation}
	\sigma_u(x)=\sigma_{I'}(x)\big|_{I'=tu}.
	\label{eq:kva-sigma-identification}
\end{equation}
This is the precise reason the local measure in the theorem is the same $\sigma_I$ introduced independently from the Jacobi symbol in section \ref{sec:canonical-pair}. Writing the limiting zero density as $\rhozero(x;\xi)$, the theorem becomes
\begin{equation}
	\rhozero(x;\xi)=\frac1\xi\int_0^\xi\dd u\,\sigma_u(x).
	\label{eq:kva-xi}
\end{equation}
Now introduce the dimensionful variables
\begin{equation}
	I=t\xi,
	\qquad
	I'=tu,
	\qquad
	\dd u=\frac{\dd I'}{t}.
\end{equation}
Then
\begin{align}
	\rhozero(x;I)
	&=\frac{t}{I}\int_0^{I/t}\dd u\,\sigma_u(x)
	\nonumber\\
	&=\frac1I\int_0^I\dd I'\,\sigma_{I'}(x),
\end{align}
which is \eqref{eq:zerodensity} of the main text.

\section{Weyl-Wigner conventions}
\label{app:wigner}

This appendix fixes the phase space normalization used in section \ref{sec:wigner-main}. We use $g_s$ as the semiclassical parameter playing the role ordinarily denoted by $\hbar$.

\subsection{Weyl symbol and trace}

For an operator $\widehat A$ with coordinate-space kernel $A(x_1,x_2)=\langle x_1|\widehat A|x_2\rangle$, define its Weyl symbol by \cite{Hillery:1983ms}
\begin{equation}
	A_W(x,p)=\int_{-\infty}^{\infty}\dd\eta\,
	e^{-\ii p\eta/g_s}
	A\!\left(x+\frac\eta2,x-\frac\eta2\right).
	\label{eq:app-weyl}
\end{equation}
The inverse formula is
\begin{equation}
	A\!\left(x+\frac\eta2,x-\frac\eta2\right)
	=\int\frac{\dd p}{2\pi g_s}
	e^{\ii p\eta/g_s}A_W(x,p).
	\label{eq:app-weyl-inverse}
\end{equation}
Setting $\eta=0$ and integrating over $x$ gives the trace identity
\begin{equation}
		\Tr\widehat A
		=\int\frac{\dd x\dd p}{2\pi g_s}\,A_W(x,p).
	\label{eq:app-weyl-trace}
\end{equation}
For $\widehat A=\widehat\Pi_n$, the symbol is the function $u_n(x,p)$ of \eqref{eq:wigner}, and \eqref{eq:app-weyl-trace} gives $\int\dd x\dd p\,u_n/(2\pi g_s)=n$ exactly.

\subsection{Products of operators and the Moyal product}

The Weyl symbol of an operator product is the Moyal product of the individual symbols,
\begin{equation}
	(\widehat A\widehat B)_W=A_W\star B_W,
\end{equation}
where in the present conventions
\begin{equation}
	f\star g
	=f\exp\!\left[
	\frac{\ii g_s}{2}
	\left(
	\overleftarrow{\partial_x}\overrightarrow{\partial_p}
	-\overleftarrow{\partial_p}\overrightarrow{\partial_x}
	\right)
	\right]g.
	\label{eq:app-moyal}
\end{equation}
The leading expansion is
\begin{equation}
	f\star g
	=fg+\frac{\ii g_s}{2}\{f,g\}_{x,p}+O(g_s^2),
	\qquad
	\{f,g\}_{x,p}=\partial_x f\,\partial_p g-\partial_p f\,\partial_x g.
	\label{eq:app-moyal-expansion}
\end{equation}
For the finite-rank projector,
\begin{equation}
	\widehat\Pi_n^2=\widehat\Pi_n
	\quad\Longrightarrow\quad
	u_n\star u_n=u_n.
	\label{eq:app-star-projector}
\end{equation}
In the joint limit
\begin{equation}
	g_s\to0,
	\qquad n\to\infty,
	\qquad I=ng_s\ \text{fixed},
	\label{eq:app-semiclassical-scaling}
\end{equation}
the star product reduces to the ordinary product away from rapidly varying boundary layers. Hence the leading bulk symbol satisfies
\begin{equation}
	u_I^2=u_I.
\end{equation}
The smooth bulk values are therefore $0$ and $1$, giving the familiar occupied/unoccupied phase space picture of a semiclassical spectral projector. 

\subsection{Fourier transform of the sine kernel}

The only Fourier identity needed in section \ref{sec:wigner-main} is
\begin{equation}
	\int_{-\infty}^{\infty}\frac{\dd\eta}{\pi\eta}
	e^{-\ii k\eta}\sin(A\eta)
	=\Thetaf(A-|k|),
	\qquad A>0,
	\label{eq:app-box-transform}
\end{equation}
with the usual value $1/2$ at $|k|=A$. One way to see this is to differentiate with respect to $A$:
\begin{align}
	\frac{\partial}{\partial A}
	\int\frac{\dd\eta}{\pi\eta}
	e^{-\ii k\eta}\sin(A\eta)
	&=\int\frac{\dd\eta}{\pi}
	e^{-\ii k\eta}\cos(A\eta)
	\nonumber\\
	&=\delta(k-A)+\delta(k+A),
\end{align}
and then integrate in $A$, using that the transform vanishes at $A=0$. With the bulk kernel of \eqref{eq:sinekernel}, the identifications are
\begin{equation}
	k=\frac{p}{g_s},
	\qquad
	A=\pi n\rhozero(x;I).
\end{equation}
Equation \eqref{eq:app-box-transform} therefore gives
\begin{align}
	u_I(x,p)
	&\longrightarrow
	\Thetaf\!\left(\pi n\rhozero(x;I)-\frac{|p|}{g_s}\right)
	\nonumber\\
	&=\Thetaf\!\left(\pi I\rhozero(x;I)-|p|\right),
\end{align}
which fixes the momentum boundary without any additional normalization factor.

\subsection{Area and orientation of the action integral}

If the limiting projector equals one on $\cD_I$ and zero outside, Eq.~\eqref{eq:app-weyl-trace} becomes
\begin{equation}
	n=\frac{\operatorname{Area}(\cD_I)}{2\pi g_s},
\end{equation}
so $\operatorname{Area}(\cD_I)=2\pi ng_s=2\pi I$. In terms of the upper and lower branches,
\begin{equation}
	\operatorname{Area}(\cD_I)
	=\int\dd x\,[p_+(x;I)-p_-(x;I)].
\end{equation}
With our convention $\Omega=\dd x\wedge\dd p$, a positively oriented geometric boundary is counterclockwise and obeys $\oint p\dd x=-\operatorname{Area}$. The action convention in the main text instead orients the classical contour so that $\oint p\dd x>0$; for the usual $(x,p)$ drawing this is the clockwise orientation. Thus
\begin{equation}
	I=\frac1{2\pi}\oint_{\text{action orientation}}p\,\dd x
	=\frac{\operatorname{Area}(\cD_I)}{2\pi}.
	\label{eq:app-action-orientation}
\end{equation}
The same orientation is induced by the standard $A$-cycle convention used in Sec.~\ref{sec:twocut} when the two sheets are mapped to the upper and lower real momentum branches.

\section{Bloch reduction for periodic Jacobi recursions}
\label{app:bloch}

This appendix collects the minimal Bloch-theory ingredients used in section \ref{sec:bloch-main}. We focus on the period-two recurrence of the symmetric quartic model and then briefly state the corresponding general-period result. More complete treatments of periodic Jacobi operators, Floquet theory, transfer matrices, and discriminants may be found in \cite{Teschl:2000,Fillman:2017}. The averaging of slowly varying asymptotically periodic recurrence coefficients used below is discussed in \cite{VanAssche:1999Periodic}.

\subsection{Two-site Bloch reduction}

In the symmetric two-cut phase the off-diagonal Jacobi coefficients
alternate as
\begin{equation}
	\ldots,\sqrt A,\sqrt B,\sqrt A,\sqrt B,\ldots .
\end{equation}
The frozen recurrence is therefore invariant under translation by two
Jacobi sites.  We group the sites $(2k-1,2k)$ into the $k$th unit cell
and denote their amplitudes by $(u_k,v_k)^T$.  With
$r_{2k}=A$, $r_{2k+1}=B$, and $s_n=0$, the local Jacobi equations are
\begin{align}
	xu_k
	&=
	\sqrt A\,v_k+\sqrt B\,v_{k-1},
	\nonumber\\
	xv_k
	&=
	\sqrt A\,u_k+\sqrt B\,u_{k+1}.
	\label{eq:app-cell-equations}
\end{align}
Translation by one complete cell is diagonalized by the Bloch ansatz
\begin{equation}
	\binom{u_k}{v_k}
	=
	e^{\ii kq}
	\binom{u}{v}.
\end{equation}
Equation \eqref{eq:app-cell-equations} then becomes
\begin{equation}
	x
	\binom{u}{v}
	=
	\mathcal J(q)
	\binom{u}{v},
\end{equation}
with
\begin{equation}
		\mathcal J(q)
		=
		\begin{pmatrix}
			0&
			\sqrt A+\sqrt B\,e^{-\ii q}
			\\[1mm]
			\sqrt A+\sqrt B\,e^{\ii q}&
			0
		\end{pmatrix}.
	\label{eq:app-bloch-matrix}
\end{equation}
The characteristic equation is
\begin{equation}
	x^2
	=
	A+B+2\sqrt{AB}\cos q.
	\label{eq:app-period2-dispersion}
\end{equation}
Using
\begin{equation}
	A+B=\frac{m}{g},
	\qquad
	AB=\frac{I}{g},
\end{equation}
gives the two Bloch bands quoted in \eqref{eq:bands}.

The Bloch angle $q$ translates an entire two-site cell and is therefore
conjugate to the cell index $k$, rather than to the original recursion
site $n$. Defining
\begin{equation}
	I_{\rm c}=g_sk,
\end{equation}
the same shift-algebra argument used in the one-cut problem gives
\begin{equation}
		\{q,I_{\rm c}\}=1.
	\label{eq:app-cell-canonical}
\end{equation}
Since $n=2k+O(1)$,
\begin{equation}
	I=ng_s
	=
	2I_{\rm c}+O(g_s),
\end{equation}
and hence, at leading order,
\begin{equation}
		I_{\rm c}=\frac{I}{2}.
	\label{eq:app-cellaction}
\end{equation}
This is the origin of the factor of two appearing in the canonical variables of the two-cut problem.

\subsection{Band edges and density of states}

Writing
\begin{equation}
	C=\frac{m}{g},
	\qquad
	D=2\sqrt{\frac{I}{g}},
\end{equation}
the positive Bloch band is
\begin{equation}
	x(q)=\sqrt{C+D\cos q}.
\end{equation}
Its extrema occur at $\cos q=\pm1$, giving
\begin{equation}
	a^2(I)=C-D,
	\qquad
	b^2(I)=C+D.
\end{equation}
Together with the negative eigenvalue, the local spectrum is therefore
\begin{equation}
	[-b(I),-a(I)]
	\cup
	[a(I),b(I)].
\end{equation}

The local density of states must be normalized per original Jacobi
site. Since a unit cell contains two sites and two Bloch eigenvalues,
\begin{equation}
	\sigma_I(x)
	=
	\frac12
	\sum_{\lambda=\pm}
	\int_0^{2\pi}
	\frac{\dd q}{2\pi}\,
	\delta\!\left(x-x_\lambda(q)\right).
	\label{eq:app-period2-dos-def}
\end{equation}
Differentiating
\begin{equation}
	x^2
	=
	\frac{m}{g}
	+
	2\sqrt{\frac{I}{g}}\cos q
\end{equation}
gives
\begin{equation}
	\left|
	\frac{\partial q}{\partial x}
	\right|
	=
	\frac{
		2|x|
	}{
		\sqrt{
			4I/g-\left(x^2-m/g\right)^2
		}
	}.
\end{equation}
Using the two values of $q$ associated with a point in the interior of
a band, \eqref{eq:app-period2-dos-def} becomes
\begin{equation}
		\sigma_I(x)
		=
		\frac{|x|}
		{\pi
			\sqrt{
				4I/g-\left(x^2-m/g\right)^2
		}},
		\qquad
		a(I)<|x|<b(I).
	\label{eq:app-period2-dos}
\end{equation}
The density is normalized over the two bands together,
\begin{equation}
	\int_{\mathcal B_I}\sigma_I(x)\,\dd x=1,
\end{equation}
and equivalently
\begin{equation}
	\left|
	\frac{\partial q}{\partial x}
	\right|
	=
	2\pi\sigma_I(x).
	\label{eq:app-qxdos}
\end{equation}
After choosing an orientation on each branch this gives \eqref{eq:qxdos}. When the period-two coefficients vary slowly with the recursion level, the limiting zero distribution is obtained by averaging the local density of states over the original slow variable $I$
\cite{VanAssche:1999Periodic},
\begin{equation}
		\rhozero(x;I)
		=
		\frac1I
		\int_0^I
		\dd I'\,
		\sigma_{I'}(x).
	\label{eq:app-period2-zero-average}
\end{equation}
The factor associated with the two-site cell is already included in
the normalization of $\sigma_I$. So no additional factor of two appears
in this average. Performing the integral for the quartic model gives \eqref{eq:twozero}, and hence
\begin{equation}
	p_\pm(x;I)
	=
	\pm\pi I\rhozero(x;I),
\end{equation}
as used in the main text.

\subsection{General periodic recurrences}

For completeness, consider a frozen Jacobi recurrence with period $\mr$.
Grouping $\mr$ sites into one unit cell leads either to an
$\mr\times \mr$ Bloch problem or, equivalently, to the usual transfer-matrix
description of a periodic Jacobi operator \cite{Teschl:2000,Fillman:2017}.  Let
$\mathcal M_\mr(x)$ denote the transfer matrix across one complete cell.
Its determinant is unity, and we use the full-trace convention for the
Floquet discriminant,
\begin{equation}
		\Delta_\mr(x)
		\equiv
		\Tr\mathcal M_\mr(x).
\end{equation}
The two Floquet multipliers are $e^{\pm\ii q_{\rm B}}$, so
\begin{equation}
		\Delta_\mr(x)
		=
		2\cos q_{\rm B}.
	\label{eq:app-discriminant}
\end{equation}
Hence real Bloch phases, and therefore allowed spectral bands, occur
for
\begin{equation}
	|\Delta_\mr(x)|\leq2.
\end{equation}
Differentiating the Bloch relation gives the density of states per
original Jacobi site,
\begin{equation}
		\sigma(x)
		=
		\frac{1}{\mr\pi}
		\frac{
			|\Delta_\mr'(x)|
		}{
			\sqrt{4-\Delta_\mr^2(x)}
		},
		\qquad
		|\Delta_\mr(x)|<2.
	\label{eq:app-general-periodic-dos}
\end{equation}
The factor $1/\mr$ reflects the fact that one Bloch cell contains $\mr$
original recursion sites.

The action conjugate to the Bloch angle is correspondingly the
$\mr$-site cell action
\begin{equation}
		I_{\rm c}^{(\mr)}
		=
		\frac{I}{\mr}
		+
		O(g_s),
		\qquad
		\{q_{\rm B},I_{\rm c}^{(\mr)}\}=1.
	\label{eq:app-general-cell-action}
\end{equation}
For $\mr=1$ this reduces to the one-cut pair $(\theta,I)$, while for
$\mr=2$ it gives the two-cut pair $(q,I_{\rm c})$ discussed above.

\bibliographystyle{jhep}
\bibliography{references}

\end{document}